\documentstyle[12pt,fleqn,epsfig]{article}

\catcode`\@=11
\@addtoreset{equation}{section}
\def\theequation{\arabic{section}.\arabic{equation}}
\newcommand{\beq}{\begin{equation}}
\newcommand{\eeq}{\end{equation}}

\def\tE{\tilde{E}}
\def\tB{\tilde{B}}

\def\tn{\tilde{n}}
\def \ut#1{\rlap{\lower1ex\hbox{$\sim$}}#1{}}
\def \UT#1{\rlap{\lower1ex\hbox{\scriptsize$\sim$}}#1{}}
\newcommand{\UI}[1]{^{\mbox{ } #1}}
\def\tN{\ut{N}}
\def\tM{\ut{M}}
\def\ep{\epsilon}
\def\otep{\tilde{\epsilon}}
\def\utep{\UT{\epsilon}}
\def\AD{\tilde{\Delta}}
\def\Hp{(\hat{\cal H}^{\prime}_{x_{0}})}
\newsavebox{\LBRA}
\savebox{\LBRA}(5,30){%
\begin{picture}(5,30)
\put(2,2){\thicklines\line(0,1){26}}
\put(3,2){\thicklines\oval(2,2)[bl]}
\put(3,28){\thicklines\oval(2,2)[tl]}
\end{picture}}
\newsavebox{\RBRA}
\savebox{\RBRA}(5,30){%
\begin{picture}(5,30)
\put(3,2){\thicklines\line(0,1){26}}
\put(2,2){\thicklines\oval(2,2)[br]}
\put(2,28){\thicklines\oval(2,2)[tr]}
\end{picture}}
\begin{document}
\begin{flushright}
OU-HET/217\\gr-qc/9506043\\June 1995
\end{flushright}
\vspace{0.5in}
\begin{center}\Large{\bf Combinatorial solutions to the
Hamiltonian constraint in (2+1)-dimensional Ashtekar gravity}\\
\vspace{1cm}\renewcommand{\thefootnote}{\fnsymbol{footnote}}
\normalsize\ Kiyoshi Ezawa\footnote[1]
{e-mail address: ezawa@funpth.phys.sci.osaka-u.ac.jp}
        \setcounter{footnote}{0}
\vspace{0.5in}

        Department of Physics \\
        Osaka University, Toyonaka, Osaka 560, Japan\\
\vspace{0.1in}
\end{center}
\vspace{1.2in}
\baselineskip 17pt
\begin{abstract}

Dirac's quantization of the (2+1)-dimensional analog of
Ashtekar's approach to
quantum gravity is investigated.
After providing a diffeomorphism-invariant
regularization of the Hamiltonian constraint,
we find a set of solutions
to this Hamiltonian constraint
which is a generalization of the solution
discovered by Jacobson and Smolin.
These solutions are given by particular
linear combinations of the spin-network states.
While the classical
counterparts of these solutions have degenerate metric,
due to a \lq quantum effect'
the area operator has nonvanishing action on these states.
We also discuss how to extend our results to (3+1)-dimensions.
\end{abstract}
\newpage
%%%%%%%%%%%%%%%%%%%%%
%%%%%%%%%%%%%%%%%%%%%

\baselineskip 20pt
\section{Introduction}

About a decade ago Ashtekar discovered the new canonical variables
which describe the canonical formulation of general relativity
\cite{ashte}. These new variables simplify the form of the
Hamiltonian constraint compared to the ADM formalism\cite{ADM}.
Thus we expect that, using these new variables,
we can solve the Wheeler-Dewitt equation\cite{dewitt},
namely the quantum version of the Hamiltonian constraint equation,
whose solution has not yet been found
in the conventional metric formulation.
Moreover, because Ashtekar's new variables can be regarded as an
$SL(2,{\bf C})$-connection and its conjugate momentum,
we can embed the phase space of general relativity
into that of an $SL(2,{\bf C})$ gauge theory\cite{ashte}.
These virtues of the new variables have led many people
to vigorous studies\cite{schil} on
the nonperturbative formulation of quantum general gravity
in terms of the new variables, namely,
Ashtekar's formalism.

While much progress has been made on Ashtekar's formalism, we have
not yet reached the complete formulation of nonperturbative
quantum general relativity because we have not yet overcome
several problems. We list a few of these problems:\\
i) The problem
of constructing the physical Hilbert space.
We can separate this problem into
two parts. One is that of finding all the
solutions to the constraint equations
which involves solving the Wheeler-Dewitt equation.
The other is that
of constructing inner product in the space of
the solutions to all the
constraints; the problem of imposing the \lq reality condition'
is considered to be contained in this problem.\\
ii) The issue of constraint closure under the commutator algebra.
This is intimately related to the problem of choosing an appropriate
operator ordering and that of inventing a regularization which is
physically relevant.

Under such situation it would be useful to study some toy models
which give some lessons on the technical and conceptual problems
in the full (3+1)-dimensional theory.
(2+1)-dimensional general relativity is considered as
one of such toy models\cite{loll}.

It has been shown by Achucarro and Townsend\cite{achu}
and Witten\cite{witt}
that the first order form of (2+1)-dimensional

Einstein gravity is equivalent to the Chern-Simons gauge theory
with a non-compact gauge group $G$, where $G$ is $SO(3,1)$,
$ISO(2,1)$, and $SO(2,2)$ when the cosmological constant is
positive, zero, and negative respectively.
Witten further showed that the
phase space of the canonical formulation of this
(2+1)-dimensional Einstein gravity is described by
the moduli space of flat
$G$-connections on a 2 dimensional space $\Sigma$
modulo gauge transformations\cite{witt}, which is finite dimensional.
We will refer to this canonical formalism as Witten's formalism.
Because Witten's formalism has a form analogous to
that of Ashtekar's formalism, it has also been investigated
as a toy model of Ashtekar's formalism\cite{husa}.
However, there is an essential difference between
these two formulations
\footnote{In this paper we consider the cosmological
constant to be zero, except in \S 2.}.
While the constraints in Witten's formalism is at most
first order in conjugate momenta, those in Ashtekar's formalism
involves the Hamiltonian constraint
which is second order in conjugate momenta.
The reason that this difference is essential is the following.
If all the constraints
are at most first order in momenta,
Dirac's quantization\cite{dirac} and
the reduced phase space (RPS) quantization\cite{fadeev},
namely the quantization on the
space of all the classical solutions to the constraints,
yield almost the same results. This is not the case
when there exist constraints
which are more than first order in momenta;
in this case the Hilbert space
of Dirac's quantization is expected to
be larger than that of the RPS quantization\cite{ashte2}.
In (3+1)-dimensions we cannot in practice carry out the RPS
quantization, because this quantization requires the full knowledge
of the solutions to Einstein equations.

We therefore suspect that, while Witten's formalism may be useful
in obtaining conceptual intuition, the techniques
developed there cannot be applied to Ashtekar's formalism.
Thus we would be glad if there exists

any formalism in (2+1)-dimensions which is more similar to
Ashtekar's formalism.

Such a formulation indeed exists.
It was found by Bengtsson\cite{beng}
in the context of reformulating the
constraint algebra of quantum gravity
from the Yang-Mills fields.
The relevant gauge group is $SO(2,1)$, or
$SL(2,{\bf R})$, which is isomorphic to the local Lorentz group in
(2+1)-dimensions. In contrast to Witten's formalism, this
\lq (2+1)-dimensional Ashtekar formalism' seems to be
investigated scarcely, except in a few works\cite{mano}\cite{bren}.
{}From the reasoning mentioned above, however, studying this
formalism is expected to yield many useful intuitions
on the quantum gravity in (3+1)-dimensions,
both in the technical and the conceptual aspects.

In this paper we investigate the quantization of
this (2+1)-dimensional Ashtekar formalism.
We will adopt Dirac's quantization procedure, in the representation
where the spin connection is diagonalized. We use
spin network states translated in terms of the spinor representation.
A spin network state\cite{penrose} is a generalization of the
Wilson loop to the \lq graphs', i.e. the set of curves

embedded in the spatial hypersurface $\Sigma$.
It is known\cite{baez} that the spin network states
form an orthogonal basis in the space $L^{2}({\cal A}/{\cal G})$
of gauge-invariant square-integrable functionals of connections, at
least when the gauge group G is compact.
Returning to (2+1)-dimensional Ashtekar formalism,
the Hamiltonian constraint operator needs to be regularized because
it involves two functional derivatives at a point.
We introduce a regularization
which preserves the covariance under the
diffeomorphisms. This regularization is a sort of point-splitting
regularization. The difference from the conventional ones is
that we use the curvilinear coordinate frame in which
the curves in the \lq graph' serve as coordinate curves.
A merit of this regularization is that we can get rid of
\lq acceleration terms'\cite{bren} which manifestly violates the
diffeomorphism covariance. After the regularization
we work out the action of the Hamiltonian constraint
on the \lq basic configurations', each of which consists of one or
two spinor parallel propagators. Then the evaluation of the
action of the Hamiltonian constraint on the spin network states
reduces to the problem of the combinatorics.
This simplifies the attempts to look for solutions
to the Hamiltonian constraint equation.
We will construct in this paper \lq combinatorial solutions',
in each of which the action of the Hamiltonian constraint
on the ingredients cancels in an elementary algebraic way.
The set of these solutions is a generalization
of the solution found by Jacobson and
Smolin\cite{jacob} to spin network
states. Each solution is labelled by
a set $\{\alpha_{i}\}$ of smooth loops
each of which is equipped with the
spin-$\frac{l_{i}}{2}$ representation
of $SL(2,{\bf R})$. We will see that classical
counterparts of these solutions
are solutions to the classical Hamiltonian
constraint having degenerate
metric. It is shown, however, these solutions can have nonzero area
due to a sort of quantum effect. In this paper we do not use
the essential merit of (2+1)-gravity, i.e. its topological nature.
Most of the results obtained in this paper can therefore be
extended to (3+1)-dimensions.
In particular, the combinatorial solutions remain
to be the solutions to
(3+1)-dimensional Hamiltonian constraint
if we appropriately modify the intertwining operators.

The organization of this paper is as follows.
After briefly reviewing (2+1)-dimensional Ashtekar formalism
and the spin network states in \S 2,
we provide a diffeomorphism-covariant
regularization in \S 3.
\S 4 is the main part of this paper. There we
provide a set of combinatorial solutions as well as the solutions
which have been known up to now.
In \S 5 we investigate the action of a few operators which measure
e.g. the area and the length, on the combinatorial solutions.
Finally in \S 6, after briefly summarizing the obtained results,
we make an attempt to extend these results to (3+1)-dimensions.
In this paper we use graphical representations frequently.
In Appendix we list the action of the Hamiltonian constraint
on the basic configurations in terms of the graphical representation.

Here we give the convention for the
indices and the signatures of the metric
used in this paper: i)$\mu,\nu,\rho,\cdots$
denote 2+1 dimensional spacetime
indices and the metric $g_{\mu\nu}$ has the signature $(-,+,+)$;
ii) $i,j,k,\cdots$ are  used for spatial indices;
iii) $a,b,c,\cdots$ represent indices of the $SO(2,1)$ representation
of the local Lorentz group, with the metric
$\eta_{ab}={\rm diag}(-,+,+)$;
iv) $A,B,C,\cdots$ are indices of the
$SL(2,{\bf R})$ spinor representation;
v) $\epsilon_{abc}$ stands for the totally antisymmetric
pseudo-tensor with $\epsilon_{012}=-\epsilon^{012}=1$;
vi) $\otep^{ij}$ ($\utep_{ij}$) denotes the totally
antisymmetric tensor density of weight $+1$ ($-1$) with
$\otep^{12}=\utep_{12}=1$;
and vii) $\epsilon^{AB}$ and $\epsilon_{AB}$ are
the rank-2 antisymmetric
tensors with $\epsilon^{12}=\epsilon_{12}=1$.

%%%%%%%%%%%%%%%%%%%%%%%%%%%%%%%%%%%%%%%%%%%%%%%%%%%%%%%%%%%%%%%%%%%

\section{Preliminaries }

In this section we provide two backgrounds which are necessary for
reading this paper, namely (2+1)-dimensional Ashtekar formalism
and the spin-network states.

\subsection{Ashtekar's formalism in (2+1)-dimensions}

(2+1)-dimensional analog of Ashtekar's approach to quantum
general relativity was first discovered
by Bengtsson\cite{beng} in the context
of the Poisson algebra of the constraints
constructed from the Yang-Mills
fields. Here we will briefly look at this
\lq(2+1)-dimensional Ashtekar
formalism' from the viewpoint of the Lagrangian formalism.

In the first order formalism of
(2+1)-dimensional general relativity, we use as
independent variables the triad one-form
$e^{a}=e_{\mu}^{a}dx^{\mu}$ and the
spin connection $\omega^{ab}=\omega_{\mu}^{ab}dx^{\mu}$.
Its action is given by the Einstein-Palatini action possibly
with a cosmological constant $\Lambda$:
\begin{equation}
I_{EP} = \int_{M}\epsilon_{abc}e^{a}\wedge
        [d\omega^{bc}+\omega^{b}_{\mbox{ }d}\wedge\omega^{dc}-
		\frac{1}{3}\Lambda e^{b}\wedge e^{c}].
\end{equation}
Let us assume $M\approx R\times\Sigma$ and
construct a canonical formalism.
For simplicity we restrict ourselves to the case
when the spatial hypersurface
$\Sigma$ is a compact, oriented two
dimensional manifold without boundary.
If we set
$\omega^{a}\equiv\frac{1}{2}\epsilon^{a}_{\mbox{ }bc}\omega^{bc}$
and make a naive (2+1)-decomposition: $x^{0}=t$,
we obtain the canonical formulation a la Witten\cite{witt}:
\begin{equation}
I_{W}=(I_{EP})_{|M\approx R\times\Sigma}
=\int dt\int_{\Sigma}d^{2}x
(-2\otep^{ij}e_{ia}\dot{\omega}_{j}^{a}+e_{t}^{a}\Psi_{a}+
\omega_{ta}{\cal G}^{a}). \label{eq:Wac}
\end{equation}
The first order constraints obtained from the variations of
the Lagrange multipliers $e_{t}^{a}$ and $\omega_{ta}$ are:
\begin{eqnarray}
\Psi^{a} &=&\otep^{ij}(F^{a}_{ij}-\Lambda
\epsilon^{a}_{\mbox{ }bc}e_{i}^{b}e_{j}^{c}), \qquad
F_{ij}^{a}\equiv\partial_{i}
\omega_{j}^{a}-\partial_{j}\omega_{i}^{a}+
\epsilon^{a}_{\mbox{ }bc}\omega_{i}^{b}\omega_{j}^{c},
\nonumber \\*
{\cal G}^{a} &=& 2\otep^{ij}(\partial_{i}e_{j}^{a}
+\epsilon^{a}_{\mbox{ }bc}\omega_{i}^{b}e_{j}^{c}).
\label{eq:Wcon}
\end{eqnarray}
It is well known that
the physical phase space of this formalism
is isomorphic to the moduli space
of flat $G$-connections on $\Sigma$ modulo gauge
transformations. The gauge
group $G$ is $SO(3,1)$, $ISO(2,1)$ and $SO(2,2)$
when the cosmological constant
$\Lambda$ is positive, zero, and negative respectively.

In order to obtain (2+1)-dimensional Ashtekar formalism, we first
remember the ADM decomposition\cite{ADM},
which is written in the first order formalism by:
$$
e_{i}^{a}e_{ja}=h_{ij},\quad e_{t}^{a}e_{ia}=N^{j}h_{ij}\equiv N_{i},
\quad e_{t}^{a}e_{ta}=-N^{2}+N^{i}N_{i},
$$
where $h_{ij}$ is the induced metric on $\Sigma$,
$N^{i}$ and $N$ are the
celebrated shift vector and the lapse function
respectively. A way to satisfy
these equations is to set
\footnote{The upper(lower) tilde stands for the
density weight $+1$
($-1$)} :
\begin{eqnarray}
e_{t}^{a}&=&N^{i}e_{i}^{a}+\tN\tn^{a} \nonumber \\*
(\tN&\equiv&\bigl(\det(h_{ij})\bigr)^{-1/2}N,\quad
\tn^{a}\equiv\frac{1}{2}\epsilon^{a}_{\mbox{ }bc}\otep^{kl}
e_{k}^{b}e_{l}^{c}).
\end{eqnarray}
By introducing the \lq momentum variable' $\tE^{i}_{a}\equiv
\otep^{ij}e_{ja}$
and a new Lagrange multiplier $\eta^{a}\equiv
-(\omega_{t}^{a}-N^{j}\omega_{j}^{a})$, we find the action of
(2+1)-dimensional analog of Ashtekar's formalism:
\begin{equation}
I_{A}=\int dt\left[\int_{\Sigma}d^{2}x
2\tE^{i}_{a}\dot{\omega}_{i}^{a}
-H \right].\label{eq:Aac}
\end{equation}
The Hamiltonian $H$ is a linear combination of
the first order constraints
which we call the Hamiltonian constraint,
the diffeomorphism constraint,
and the Gauss law constraint respectively:
\begin{eqnarray}
&&H={\cal H}(\tN)+{\cal D}_{i}(N^{i})+{\cal G}_{a}(\eta^{a}),
\nonumber \\*
&&\left\{\begin{array}{lll}
{\cal H}(\tN)&\equiv&-\int_{\Sigma}d^{2}x\tN[\epsilon_{abc}
\tE^{ib}\tE^{jc}F^{a}_{ij}-\Lambda\utep_{ik}\utep_{jl}
\tE^{ia}\tE^{jb}\tE^{l}_{a}\tE^{k}_{b}]\\
{\cal D}_{i}(N^{i})&\equiv&2\int_{\Sigma}d^{2}x
N^{i}(\tE^{j}_{a}F^{a}_{ij}-\omega_{i}^{a}D_{j}\tE^{j}_{a})=
2\int_{\Sigma}d^{2}x\tE^{j}_{a}{\cal L}_{\vec{N}}\omega_{j}^{a}\\
{\cal G}_{a}(\eta^{a})&\equiv&
2\int_{\Sigma}d^{2}x\eta^{a}D_{j}\tE^{j}_{a},
\end{array}\right.\label{eq:Acon}
\end{eqnarray}
where ${\cal L}_{\vec{N}}$ and $D_{j}$ denote
respectively the Lie derivative
w.r.t. $N^{i}$ and the covariant derivative defined by
the spin connection $\omega_{j}^{a}$.

{}From the action (\ref{eq:Aac}) we can read off
the basic Poisson brackets:
\beq
\{\omega_{i}^{a}(x),\tE^{j}_{b}(y)\}=\frac{1}{2}
\delta^{a}_{b}\delta^{j}_{i}\delta^{2}(x,y).\label{eq:bPb}
\eeq
Under the Poisson bracket, the diffeomorphism and the
Gauss law constraints respectively generate (infinitesimal) spatial
diffeomorphisms
and the local Lorentz transformations:
\begin{eqnarray}
\{(\omega_{i}^{a},\tE^{i}_{a}),{\cal D}_{i}(N^{i})\}&=&
({\cal L}_{\vec{N}}\omega_{i}^{a},\quad
{\cal L}_{\vec{N}}\tE^{i}_{a}) \nonumber \\*
\{(\omega_{i}^{a},\tE^{i}_{a}),{\cal G}_{a}(\eta^{a})\}&=&
(-D_{i}\eta^{a},\quad -\ep_{abc}\tE^{ib}\eta^{c}).
\end{eqnarray}
The Hamiltonian constraint generates the \lq bubble-time evolutions':
\begin{eqnarray}
\{\omega_{i}^{a},{\cal H}(\tN)\}&=&-\tN(\ep^{a}_{\mbox{ }bc})\tN^{jb}
F^{c}_{ij}-2\Lambda\ep_{ik}\ep_{jl}\tE^{la}\tE^{jb}\tE^{k}_{b})
\nonumber \\*
\{\tE_{a}^{i},{\cal H}(\tN)\}&=&D_{j}(\tN\ep_{abc}\tE^{ib}\tE^{jb}).
\label{eq:diflL}
\end{eqnarray}
Using these equations, we can compute the Poisson brackets between
the constraints\cite{beng}:
\begin{eqnarray}
\{{\cal D}_{i}(M^{i}),{\cal D}_{j}(N^{j})\}&=&-{\cal D}_{i}
({\cal L}_{\vec{N}M^{i}}),\nonumber \\*
\{{\cal G}_{a}(\eta^{a}),{\cal D}_{j}(N^{j})\}&=&-{\cal G}_{a}
({\cal L}_{\vec{N}}\eta^{a}),\nonumber \\*
\{{\cal H}(\tN),{\cal D}_{j}(N^{j})\}&=&
-{\cal H}({\cal L}_{\vec{N}}\tN),
\label{eq:CPA}\\
\{{\cal G}_{a}(\eta^{a}),{\cal G}_{b}(\eta^{\prime a})\}
&=&{\cal G}_{a}
(\ep^{abc}\eta_{b}\eta^{\prime}_{c}), \nonumber \\*
\{{\cal H}(\tN),{\cal G}_{a}(\eta^{a})\}&=&0, \nonumber \\*
\{{\cal H}(\tN),{\cal H}(\tM)\}&=&{\cal D}_{i}(K^{i})+{\cal G}_{a}
(\omega^{a}_{i}K^{i}),\quad K^{i}\equiv \tE^{ib}\tE^{j}_{b}
(\tN\partial_{j}\tM-\tM\partial_{j}\tN).\nonumber
\end{eqnarray}

Here we will make some remarks. The Witten constraint equations:
$$
\Psi^{a}\approx{\cal G}^{a}\approx0
$$
and the Ashtekar constraint equations:
\beq
{\cal H}=-\tn^{a}\Psi_{a}\approx0,\quad
{\cal V}_{i}\equiv -e_{i}^{a}\Psi_{a}\approx0,\quad
{\cal G}^{a}\approx 0
\eeq
are equivalent if $e_{i}^{a}$ ($i=1,2$) and
$\tn^{a}$ form a non-degenerate frame, namely,
if the space-time metric
$g_{\mu\nu}=e^{a}_{\mu}e_{\nu a}$ is non-degenerate.
If $g_{\mu\nu}$ is
degenerate, however, there exist solutions to
the Ashtekar constraint equations
which are not subject to the Witten constraint equations.
The phase space of (2+1)-Ashtekar formalism is thus expected to
contain that of Witten's formalism as a subspace\cite{mano}.

%%%%%%%%%%%%%%%%%%%%%%%%%%%%%%%%%%%%%%%%%%%%%%%%%%%%%%%%%%%%%%%%%%%%

\subsection{Spin network states}

In this paper we will make an attempt to quantize (2+1)-dimensional
gravity on the space ${\cal A}$ of spin connections
$\omega_{i}^{a}$, which can be regarded as
$SL(2,{\bf R})$-connections.
As will be seen in the next section,
the Gauss law constraint tells us
that the wavefunctions be the gauge invariant functionals of the spin
connection. We are thus interested
in the functions on the quotient space
${\cal A}/{\cal G}$, where ${\cal G}$ is
the group of $SL(2,{\bf R})$ gauge
transformations.

It is well known that the Wilson loops
yield the gauge invariant information
on the connection\cite{gile}. A Wilson loop is defined if a loop
$\gamma:[0,1]\rightarrow\Sigma$
($\gamma(0)=\gamma(1)$) and a representation
$\rho$ of $SL(2,{\bf R})$ is given:
\beq
W(\gamma,\rho)\equiv{\rm Tr}\rho(h_{\gamma}[0,1]),\label{eq:wilson}
\eeq
where
\beq
h_{\alpha}[0,1]\equiv {\cal P}\exp\{\int_{0}^{1}ds\dot{\alpha}^{i}(s)
A_{i}(\alpha(s))\} \label{eq:holo}
\eeq
is the parallel propagator of the
$SL(2,{\bf R})$ connection
$A_{i}\equiv\omega^{a}_{i}J_{a}$ ($J_{a}$: the
generators of $SL(2,{\bf R})$) along a curve
$\alpha:[0,1]\rightarrow\Sigma$,
where $\alpha(0)$ does not in general coincide with $\alpha(1)$.
Under the gauge transformation
$$
A_{i}(x)\rightarrow g(x)A_{i}(x)g^{-1}(x)
-\partial_{i}g(x)g^{-1}(x)\quad  \left(g(x)\in SL(2,{\bf R})\right),
$$
the parallel propagator $h_{\alpha}[0,1]$ transforms as:
\beq
h_{\alpha}[0,1]\rightarrow
g(\alpha(0))h_{\alpha}[0,1]g^{-1}(\alpha(1)).
\label{eq:holotr}
\eeq
Gauge invariance of the Wilson loop (\ref{eq:wilson})
is an immediate consequence of this.

Spin network states are the generalized version of this Wilson
loop\cite{penrose}\cite{baez}.
A spin network state is in one-to-one correspondence with
a \lq colored graph'. A colored graph is
specified by a set of \lq colored edges', the edges $e$
(the segments of the curves in $\Sigma$) each of
which being labelled by a representation
$\rho_{e}$of $SL(2,{\bf R})$,
and a set of \lq colored vertices', the vertices $v$
each of which being equipped
with an intertwining operator $i_{v}$
\footnote{The intertwining operator $i_{v}$ is an operator
which extracts the trivial representation from the direct product
of the representations labelling
the edges which have an end at $v$.}.
The representations $\rho_{e}$, of course, must satisfy some
conditions at each vertex in order to
ensure the gauge invariance.

The $SL(2,{\bf R})$ group has finite
dimensional representations and
infinite dimensional ones.
We will henceforth restrict our attention to
the former. It is a well-known fact that the
($l+1$)-dimensional irreducible
representation, i.e. the spin-$\frac{l}{2}$ representation,
is expressed as the symmetrized tensor product of $l$
copies of the fundamental representation
(the spinor representation).
Thus we can make our analysis by using
the spinor representation only.

The parallel propagator in the spinor
representation along a curve $\alpha$ is
\beq
h_{\alpha}[0,1]_{A}\UI{B}=
\left[{\cal P}\exp\int_{0}^{1}ds\dot{\alpha}^{i}(s)
\omega^{a}_{i}(\alpha(s))\lambda_{a}\right]_{A}\UI{B},
\eeq
where $\lambda_{a}$ are spin-$\frac{1}{2}$ generators subject to
$\lambda_{a}\lambda_{b}=
\frac{1}{4}\eta_{ab}+\frac{1}{2}\ep_{abc}\lambda^{c}$.
This spinor propagator is subject to the following identity:
\beq
\ep_{AD}\ep^{BC}h_{\alpha}[0,1]_{C}\UI{D}=
(h_{\alpha}[0,1]^{-1})_{A}\UI{B}
=h_{\alpha^{-1}}[0,1]_{A}\UI{B},\label{eq:identity1}
\eeq
where $\alpha^{-1}$: $\alpha^{-1}(s)=\alpha(1-s)$
is the curve obtained by
reversing the orientation of $\alpha$.
This identity is useful because it
enables us to choose a relevant orientation
of the propagators according to
a particular problem.

In the spinor representation, there are three
invariant tensors (tensors which
are invariant under the $SL(2,{\bf R})$ transformations)
$\ep_{B}\UI{A}\equiv\delta^{A}_{B}$,
$\ep^{AB}$ and $\ep_{AB}$. Because the intertwining operators
are constructed from the invariant tensors,
the total rank of an intertwining
operator is even.
Thus the sum of the numbers of the spinor propagators
ending and starting at a  vertex is necessarily even. Using identity
(\ref{eq:identity1}) and the equation
$\ep_{AC}\ep^{BC}=\delta^{B}_{A}$,
we can equalize at each vertex
the number of in-coming propagators
with that of out-going propagators.
As a consequence we can regard the intertwining operators to be a
sum of the products of $\delta_{A}^{B}$.
The product of the propagators along

two successive curves $\alpha$, $\beta$
($\alpha(1)=\beta(0)$) is the
propagator along the composite curve $\alpha\circ\beta$:
$$
h_{\alpha}[0,1]_{A}\UI{C}h_{\beta}[0,1]_{C}\UI{B}
=h_{\alpha\circ\beta}[0,1]_{A}\UI{B}.
$$
We can thus consider the spin network states to be
(the linear combinations of)
products of the Wilson loops in the spinor
representation along the loops
each of which is composed of the edges in the graph.

In addition to eq.(\ref{eq:identity1}),
there are two useful identities
in the spinor representation; the two-spinor identity
\beq
\delta_{A}^{B}\delta_{C}^{D}-\delta_{A}^{D}\delta_{C}^{B}
=\ep_{AC}\ep^{BD}
\quad(\mbox{ or }
\phi^{A}\ep^{BC}+\phi^{B}\ep^{CA}+\phi^{C}\ep^{AB}=0)
\label{eq:2spi}
\eeq
and the Fiertz identity
\beq
(\lambda^{a})_{A}\UI{B}(\lambda_{a})_{C}\UI{D}=
\frac{1}{2}(\delta_{A}^{D}\delta_{C}^{B}
-\frac{1}{2}\delta_{A}^{B}\delta_{C}^{D}).\label{eq:fiertz}
\eeq
The two-spinor identity tells us that the antisymmetrized
tensor product of the two identical
spinor propagators gives the trivial
representation\footnote{The indices enclosed by
$[]$ and $()$ are supposed
to be antisymmetrized and symmetrized respectively.}:
\beq
h_{\alpha}[0,1]_{A}\UI{[B}h_{\alpha}[0,1]_{C}\UI{D]}=
\frac{1}{2}\ep_{AC}\ep^{BD}
\label{eq:2spi2}
\eeq
and that, at an intersection
$\alpha(s_{0})=\beta(t_{0})$ of two curves
$\alpha$ and $\beta$, the following identity holds:
\begin{eqnarray}
h_{\alpha}[0,1]_{A}\UI{B}h_{\beta}[0,1]_{C}\UI{D}-
(h_{\alpha}[0,s_{0}]h_{\beta}[t_{0},1])_{A}\UI{D}
(h_{\beta}[0,t_{0}]h_{\alpha}[s_{0},1])_{C}\UI{B} \nonumber \\*
=(h_{\alpha}[0,s_{0}]h_{\beta^{-1}}[1-t_{0},1])_{A}\UI{E}\ep_{EC}
(h_{\beta^{-1}}[0,1-t_{0}]h_{\alpha}[s_{0},1])_{F}\UI{B}\ep^{FD}.
\label{eq:2spi3}
\end{eqnarray}
Owing to eq.(\ref{eq:2spi2}) we do not have to
take account of antisymmetrizing
the identical propagators.

For the purpose of the calculation,
it is convenient to introduce the graphical representation.
We will denote a spinor propagator
$h_{\alpha}[0,1]_{A}\UI{B}$ by an arrow from $\alpha(0)$
to $\alpha(1)$ with its tail (tip) being
associated with the spinor index
$A$ ($B$). Identities (\ref{eq:identity1}),(\ref{eq:2spi2}) and
(\ref{eq:2spi3}) are then expressed as follows:

%%%%%%%%%%%%(identity1)%%%%%%%%%%%%%%%%%%%%%%%%%%%%%%%%%%%%%%%%%
\begin{figure}[ht]
\begin{picture}(150,30)
\put(0,0){\makebox(20,30){$\ep_{AD}\ep^{BC}$}}
\put(20,0){\usebox{\LBRA}}
\put(30,5){\vector(0,1){20}}
{\scriptsize
\put(30,2){$C$}
\put(30,25){$D$}}
\put(40,0){\usebox{\RBRA}}
\put(45,0){\makebox(30,30){$=$}}
\put(75,0){\usebox{\LBRA}}
\put(85,25){\vector(0,-1){20}}
{\scriptsize
\put(85,2){$B$}
\put(85,25){$A$}}
\put(95,0){\usebox{\RBRA}}
\put(100,0){\makebox(5,30){,}}
\put(120,0){\makebox(20,30){(\ref{eq:identity1}$)^{\prime}$}}
\end{picture}
\end{figure}

%%%%%%%%%%%%%%(2spi2)%%%%%%%%%%%%%%%%%%%%%%%%%%%%%%%%%%%%%%%
\begin{figure}[ht]
\begin{picture}(150,65)(-15,0)
\put(-5,35){\usebox{\LBRA}}
\put(8,40){\vector(0,1){20}}
\put(12,40){\vector(0,1){20}}
{\scriptsize
\put(8,37){$A$}
\put(8,60){$B$}
\put(12,37){$C$}
\put(12,60){$D$}}
\put(20,35){\makebox(20,30){$-$}}
\put(48,52){\vector(0,1){8}}
\put(48,40){\line(0,1){8}}
\put(52,52){\vector(0,1){8}}
\put(52,40){\line(0,1){8}}
\put(48,48){\line(1,1){4}}
\put(48,52){\line(1,-1){4}}
{\scriptsize
\put(48,37){$A$}
\put(48,60){$B$}
\put(52,37){$C$}
\put(52,60){$D$}}
\put(60,35){\usebox{\RBRA}}
\put(65,35){\makebox(30,30){$=$ $\ep_{AC}\ep^{BD}$ ,}}
\put(110,35){\makebox(20,30){(\ref{eq:2spi2}$)^{\prime}$}}

%%%%%%%%%%%%%%%%%%(2spi3)%%%%%%%%%%%%%%%%%%%%%%%%%%%%%%%%%%%%%

\put(-5,0){\usebox{\LBRA}}
\put(10,5){\vector(0,1){20}}
\put(0,15){\vector(1,0){20}}
{\scriptsize
\put(10,2){$A$}
\put(10,25){$B$}
\put(0,15){$C$}
\put(20,15){$D$}}
\put(20,0){\makebox(20,30){$-$}}
\put(49,16){\vector(0,1){9}}
\put(40,16){\line(1,0){9}}
\put(51,14){\vector(1,0){9}}
\put(51,5){\line(0,1){9}}
{\scriptsize
\put(51,2){$A$}
\put(49,25){$B$}
\put(40,16){$C$}
\put(60,14){$D$}}
\put(60,0){\usebox{\RBRA}}
\put(65,0){\makebox(10,30){$=$}}
\put(75,0){\makebox(10,30){$\ep_{EC}\ep^{FD}$}}
\put(85,0){\usebox{\LBRA}}
\put(101,16){\vector(0,1){9}}
\put(101,16){\line(1,0){9}}
\put(99,14){\vector(-1,0){9}}
\put(99,5){\line(0,1){9}}
{\scriptsize
\put(99,2){$A$}
\put(101,25){$B$}
\put(110,16){$F$}
\put(90,14){$E$}}
\put(110,0){\usebox{\RBRA}}
\put(115,0){\makebox(15,30)[l]{. (\ref{eq:2spi3}$)^{\prime}$}}
\end{picture}
\end{figure}
%%%%%%%%%%%%%%%%%%%%%%%%%%%%%%%%%%%%%%%%%%%%%%%%%%%%%%%

%%%%%%%%%%%%%%%%%%%%%%%%%%%%%%%%%%%%%%%%%%%%%%%%%%%%%%%%%%%%%%%%%%%

\section{Dirac quantization}

Now that we have provided the necessary backgrounds, let us try to
quantize (2+1)-dimensional gravity in Ashtekar's form. In this paper
we will adopt the quantization procedure proposed
by Dirac\cite{dirac}.
In Dirac's quantization procedure, we first quantize the phase space
by promoting the canonical variables
$(\omega_{i}^{a},\tE^{i}_{a})$ to
the operators $(\hat{\omega}_{i}^{a},\hat{\tE}^{i}_{a})$
which act on an
appropriately chosen Hilbert space, and
by replacing $i$-times the basic Poisson bracket
(\ref{eq:bPb}) with the quantum commutation relation:
\beq
[\hat{\omega}_{i}^{a}(x),\hat{\tE}^{j}_{b}(y)]=\frac{i}{2}
\delta^{a}_{b}\delta_{i}^{j}\delta^{2}(x,y).\label{eq:qcom}
\eeq
We will use as a pre-constrained Hilbert space
the space $L^{2}(\mu,{\cal A})$
of the square-integrable functionals of
$\omega^{a}_{i}$ w.r.t a certain
measure $\mu$ on ${\cal A}$. $\hat{\omega}_{i}^{a}$
and $\hat{\tE}^{i}_{a}$
then act on the wavefunctions by multiplication
and functional differentiation,
respectively:
\beq
\hat{\omega}_{i}^{a}(x)\cdot\Psi(\omega)=
\omega_{i}^{a}(x)\cdot\Psi(\omega),
\quad\hat{\tE}^{i}_{a}(x)\cdot\Psi(\omega)
=-\frac{i}{2}\frac{\delta}{\delta\omega_{i}^{a}(x)}\Psi(\omega).
\eeq
Next we impose the constraint equations as the operator equations
which restrict the wavefunctions allowed in the theory\footnote{%
There is the issue on the choice of
operator ordering which is intimately
related to the problem on the closure of the constraint algebra
\cite{koda}\cite{bori}. We will not discuss on these issues and
simply choose the ordering in which the momentum operators
$\hat{\tE}^{i}_{a}$
are placed to the right of the spin connections
$\hat{\omega}_{i}^{a}$.}.
The Gauss law and the diffeomorphism constraints:
\begin{eqnarray}
\hat{\cal G}_{a}(\eta^{a})\Psi(\omega)=i\int_{\Sigma}d^{2}x
D_{i}\eta^{a}(x)\frac{\delta}{\delta\omega_{i}^{a}(x)}\Psi(\omega)
\nonumber \\*
\hat{\cal D}_{i}(N^{i})\Psi(\omega)=
-i\int_{\Sigma}d^{2}x{\cal L}_{\vec{N}}
\omega^{a}_{i}(x)\frac{\delta}{\delta\omega_{i}^{a}(x)}\Psi(\omega)
\end{eqnarray}
have simple geometrical interpretations. The former requires the
wavefunctionals to be invariant under the
small gauge transformations,
and the latter tells us that the
wavefunctionals be invariant under the
spatial diffeomorphisms connected to
the identity.

Because the spin network states are shown to span the dense subset
of the space of the gauge-invariant functionals of $SL(2,{\bf R})$
connections\cite{baez}, to work in the set
of these states (or its completion)
automatically satisfies the Gauss law constraint.
Solving the diffeomorphism
constraint, on the other hand, requires
a special consideration and we will
postpone this procedure until the next section. Here we only mention
that  the exponentiated diffeomorphism
constraint gives the diffeomorphism
operator:
\beq
\hat{U}(\phi)\omega_{i}^{a}(x)=
\frac{\partial\phi^{-1}(x)^{j}}{\partial x^{i}}
\omega_{j}^{a}(\phi^{-1}(x)),
\eeq
whose action on a parallel propagator can be
cast into the action on the
curve along which the propagator is evaluated:
\beq
\hat{U}(\phi)h_{\alpha}[0,1]=h_{\phi^{-1}\cdot\alpha[0,1]}.
\label{eq:DIFF}
\eeq

Unlike the previous two constraints,
to define the Hamiltonian constraint operator
$$
\hat{\cal H}(\tN)=\frac{1}{4}\int_{\sigma}d^{2}x
\tN(x)\ep_{abc}F^{a}_{ij}
\frac{\delta}{\delta\omega_{ib}(x)}
\frac{\delta}{\delta\omega_{jc}(x)},
$$
we have to prescribe some regularization because it involves two
functional derivatives at an identical point.
The problem is that whether we
can find a regularization which preserves
the closure of the constraints
under the commutator algebra. In particular,
it is important whether there
exists a regularization which is invariant under the diffeomorphisms.
In (2+1)-dimensions we can regularize the
Hamiltonian constraint at least
in a diffeomorphism-invariant manner if we restrict the types of
graphs on which the spin network states are defined.

%%%%%%%%%%%%%%%%%%%%%%%%%%%%%%%%%%%%%%%%%%%%%%%%%%%%%%%%%%%%%%%%%%%%%

\subsection{A diffeomorphism-covariant regularization}

Among the regularizations of the Hamiltonian constraint,
the most familiar
one is the point-splitting regularization
\cite{smol}\cite{bren}\cite{bori}:
\begin{eqnarray}
\hat{\cal H}(\tN)&\!\!=\!\!&
\lim_{\ep\rightarrow0}\hat{\cal H}^{\ep}(\tN),
\label{eq:reghami} \\
\hat{\cal H}^{\ep}(\tN)&\!\!=\!\!&
\int_{\Sigma}d^{2}x\int_{\Sigma}d^{2}y
\tN(x)\tilde{f}_{\ep}(x,y)\utep_{ij}{\rm Tr}
(h_{yx}[0,1]\tB(x)\lambda^{b}h_{xy}[0,1]\lambda^{c})
\frac{\delta}{\delta\omega_{ib}(x)}
\frac{\delta}{\delta\omega_{jc}(y)},
\nonumber
\end{eqnarray}
where $h_{xy}[0,1]$ is the parallel propagator along a curve from $x$
to $y$ which shrinks to $x$ in the limit $y\rightarrow x$,
$\tB\equiv\tB^{a}\lambda_{a}\equiv\frac{1}{2}
\otep^{ij}F^{a}_{ij}\lambda_{a}$
is the magnetic field of the spin connection and the regulator
$\tilde{f}_{\ep}(x,y)$ is subject to the condition
$$
\tilde{f}_{\ep}(x,y)\stackrel{\ep\rightarrow0}{\longrightarrow}
\delta^{2}(x,y).
$$
In order to define a particular regulator,
we usually have to introduce
some extra background structures such as
a fiducial metric or a fixed frame.
These background structures do not
behave covariantly under the action
of the diffeomorphism constraint. This is the source of
the diffeomorphism non-invariance of
the regularized Hamiltonian constraint.

It is obvious that the action of $\hat{\cal H}(\tN)$ on the parallel
propagators is nonvanishing only on
the curves along which the propagators
are evaluated. Eq.(\ref{eq:DIFF})
tells us that these curves are subject to
the action of the diffeomorphism constraint.
We therefore expect that
if we introduce the curvilinear coordinate
frame in which these curves play
the role of coordinate curves (figure1),
then we can define a regularization
of the Hamiltonian constraint which
preserves the diffeomorphism covariance.
In this subsection we explicitly carry out
such a regularization procedure.

%%%%%%%%%%%% figure1 %%%%%%%%%%%%%%%%%%%%%%%%%%%
\begin{figure}[p]
\begin{center}
\epsfig{file=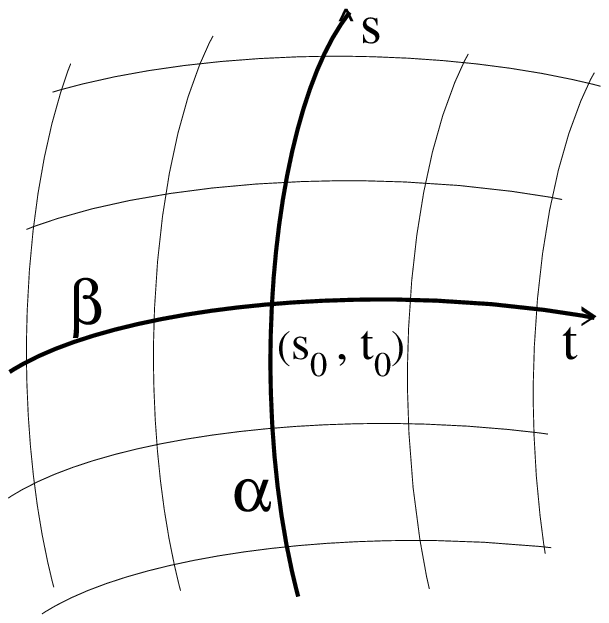,height=6cm}
\end{center}
\caption{A curvilinear coordinate
frame used in our regularization.}
\end{figure}
%%%%%%%%%%%%%%%%%%%%%%%%%%%%%%%%%%%%%%%%%%%%%%

To simplify the analysis we restrict the types of the vertices
to those at which only two smooth curves
with linear independent tangent
vectors intersect. We therefore do not consider  vertices at which
more than three independent curves intersect (figure 2(a),(b)),
cusps (figure 2(c)),or vertices at which two
independent curves with coincident tangent vectors intersect
(figure 2(d)).
We consider only configurations depicted in figure3.

%%%%%%%%%%%% figure2 %%%%%%%%%%%%%%%%%%%%%%%%%
\begin{figure}[p]
\begin{center}
\epsfig{file=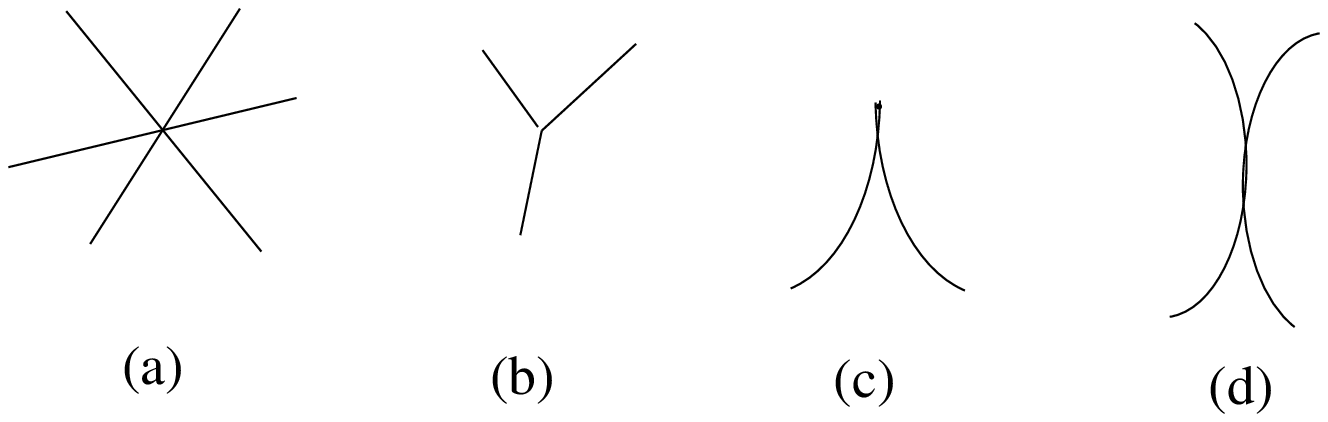,height=4cm}
\end{center}
\caption{Configurations not considered in this paper}
\end{figure}
%%%%%%%%%%%%%%%%%%%%%%%%%%%%%%%%%%%%%%%%%%%%%%%%

%%%%%%%%%%%%%%% figure3 %%%%%%%%%%%%%%%%%%%%%%%%%%%%%%%%%%%
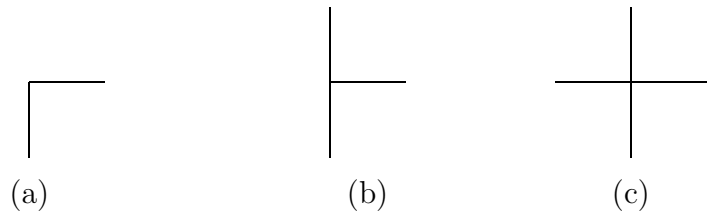
\begin{figure}[p]
\begin{center}
\begin{picture}(95,30)(0,0)

%%%%%%%%%%% fig.3(a)%%%%%%%%%%%%%%%%%%%%%%%%%%%%%

\put(10,10){\line(0,1){10}}
\put(10,20){\line(1,0){10}}
\put(0,0){\makebox(20,10){(a)}}

%%%%%%%%%%% fig.3(b)%%%%%%%%%%%%%%%%%%%%%%%%%%%%%

\put(50,10){\line(0,1){20}}
\put(50,20){\line(1,0){10}}
\put(45,0){\makebox(20,10){(b)}}

%%%%%%%%%%% fig.3(c) %%%%%%%%%%%%%%%%%%%%%%%%%%%%%

\put(90,10){\line(0,1){20}}
\put(80,20){\line(1,0){20}}
\put(80,0){\makebox(20,10){(c)}}

%%%%%%%%%%%%%%%%%%%%%%%%%%%%%%%%%%%%%%%%%%%%%%%
\end{picture}
\end{center}
\caption{The configurations considered
in this paper: (a)two-point vertex;
(b)three-point vertex; and (c)four-point vertex}
\end{figure}
%%%%%%%%%%%%%%%%%%%%%%%%%%%%%%%%%%%%%%%%%%%%%%%%%

Because the Hamiltonian constraint involves
two functional derivatives,
it is convenient to separate its action as follows:
\beq
\hat{\cal H}(\tN)=\hat{\cal H}_{1}(\tN)
+\hat{\cal H}_{2}(\tN),\label{eq:sepa}
\eeq
where $\hat{\cal H}_{1}(\tN)$ and $\hat{\cal H}_{2}(\tN)$ stand for,
respectively, the action of $\hat{\cal H}(\tN)$
on the single loops and that
on the pairs of loops. This separation simplifies
considerably the evaluation of the action of
the Hamiltonian constraint on the spin-network
states.

Now we demonstrate a few examples of the calculation.

First we consider the action of the regularized Hamiltonian
(\ref{eq:reghami})
on a single smooth loop $\alpha$:
\begin{eqnarray*}
&\!\!\!{}\!\!\!&\hat{\cal H}^{\ep}(\tN)h_{\alpha}[0,1]_{A}\UI{B}
=\int_{\Sigma}d^{2}x\int_{\Sigma}d^{2}y
\tN(x)\tilde{f}_{\ep}(x,y)\utep_{ij}{\rm Tr}
(h_{yx}[0,1]\tB(x)\lambda^{b}h_{xy}[0,1]\lambda^{c})\\
&\!\!\!{}\!\!\!&
\qquad\times\int ds\delta^{2}(x,\alpha(s))\dot{\alpha}^{i}(s)
\int dt\delta^{2}(y,\alpha(t))\dot{\alpha}^{j}(t) \\
&\!\!\!\times\!\!\!&
\left\{\theta(t-s)h_{\alpha}[0,s]\lambda_{a}h_{\alpha}[s,t]
\lambda_{b}h_{\alpha}[t,1]+
\theta(s-t)h_{\alpha}[0,t]\lambda_{b}h_{\alpha}[t,s]
\lambda_{a}h_{\alpha}[s,1]\right\}_{A}\UI{B}.\\
&\!\!\!=\!\!\!&
\int\int dsdt\utep_{ij}\dot{\alpha}^{i}(s)\dot{\alpha}^{j}(t)
\tilde{f}_{\ep}(\alpha(s),\alpha(t))\tN(\alpha(s))
{\rm Tr}(h_{ts}\tB(\alpha(s))\lambda^{b}h_{st}\lambda^{c})\\
&\!\!\!\times\!\!\!&
\left\{\theta(t-s)h_{\alpha}[0,s]\lambda_{a}h_{\alpha}[s,t]
\lambda_{b}h_{\alpha}[t,1]
+\theta(s-t)h_{\alpha}[0,t]\lambda_{b}h_{\alpha}[t,s]
\lambda_{a}h_{\alpha}[s,1]\right\}_{A}\UI{B},
\end{eqnarray*}
where we have abbreviated $h_{\alpha(s)\beta(t)}[0,1]$ to $h_{st}$.
Because $\alpha$ serves as a coordinate curve,
$\dot{\alpha}(s)\propto
\dot{\alpha}(t)$ holds even when $t$ does not
coincide with $s$ (but is
sufficiently close to $s$).
We find that the result vanishes due to the
factor $\utep_{ij}\dot{\alpha}^{i}(s)\dot{\alpha}^{j}(t)$.
Hence our
regularization does not suffer from
\lq acceleration terms'\cite{bren}
which manifestly violates diffeomorphism covariance.

Next we calculate the action on a single loop with a kink, which
we will denote by $\alpha_{1}\circ\beta_{2}$
\footnote{In this paper we usually assume
that the curves $\alpha$ and
$\beta$ intersect with each other at
$\alpha(s_{0})=\beta(t_{0})=x_{0}$.
$\alpha_{1}\subset\alpha$ is the curve
from $\alpha(0)$ to $\alpha(s_{0})$
and $\beta_{2}\subset\beta$ is the
curve from $\beta(t_{0})$ to $\beta(1)$.}:
\begin{eqnarray}
&\!\!\!{}\!\!\!&\hat{\cal H}^{\ep}(\tN)
(h_{\alpha}[0,s_{0}]h_{\beta}[t_{0},1])_{A}\UI{B}
=\hat{\cal H}^{\ep}_{1}
(\tN)(h_{\alpha}[0,s_{0}]h_{\beta}[t_{0},1])_{A}\UI{B}\nonumber \\*
&\!\!\!=\!\!\!&\int_{0}^{s_{0}}ds\int_{t_{0}}^{1}dt
\utep_{ij}\dot{\alpha}^{i}(s)\dot{\beta}^{j}(t)
\tilde{f}_{\ep}(\alpha(s),\beta(t))\nonumber \\*
&\!\!\!\times\!\!\!&\left\{\tN(\alpha(s))
{\rm Tr}(h_{ts}[0,1]\tB(\alpha(s))
\lambda^{b}h_{st}[0,1]\lambda^{c})
h_{\alpha}[0,s]\lambda_{b}h_{\alpha}[s,s_{0}]
h_{\beta}[t_{0},t]\lambda_{c}h_{\beta}[t,1]\right.\nonumber \\*
&\!\!\!-\!\!\!&\left.
\tN(\beta(t)){\rm Tr}(h_{st}[0,1]\tB(\beta(t))
\lambda^{b}h_{ts}[0,1]\lambda^{c})
h_{\alpha}[0,s]\lambda_{c}h_{\alpha}[s,s_{0}]
h_{\beta}[t_{0},t]\lambda_{b}h_{\beta}[t,1]\right\}_{A}\UI{B}
\nonumber \\*
&\!\!\!\equiv\!\!\!&\int_{0}^{s_{0}}ds\int_{t_{0}}^{1}dt
\utep_{ij}\dot{\alpha}^{i}(s)\dot{\beta}^{j}(t)
\tilde{f}_{\ep}(\alpha(s),\beta(t))\times I_{1},
\end{eqnarray}
We first simplify the integrand. By using
$h_{\alpha(s)\beta(t)}[0,1]_{A}
\UI{B}=\delta_{A}^{B}+O(\ep)$ and similar approximations, we find
\begin{eqnarray*}
I_{1}&=&\left\{\frac{\ep^{abc}}{4}\tN(x_{0})\tB_{a}(x_{0})
\left(h_{\alpha}[0,s_{0}]
(\lambda_{b}\lambda_{c}-\lambda_{c}\lambda_{b})
h_{\beta}[t_{0},1]\right)+O(\ep)\right\}_{A}\UI{B}\\
&=&-\frac{1}{2}\left\{\tN(x_{0})
h_{\alpha}[0,s_{0}]\tB(x_{0})h_{\beta}[t_{0},1]
+O(\ep)\right\}_{A}\UI{B}.
\end{eqnarray*}
The $O(\ep)$ part also transforms covariantly
under the gauge transformations.
Now we consider the part involving the
regulator:
$$
\int_{0}^{s_{0}}ds\int_{t_{0}}^{1}dt
\utep_{ij}\dot{\alpha}^{i}(s)\dot{\beta}^{j}(t)
\tilde{f}_{\ep}(\alpha(s),\beta(t)).
$$
In the conventional point splitting
regularization we fix a background
metric to define $\tilde{f}_{\ep}$. In our regularization, however,
we first exploit the
transformation property of the $\delta$-function:
$$
|\det(\partial_{i}\phi^{j}(x))|
\delta^{n}(\phi(x),\phi(y))=\delta^{n}(x,y),
$$
and we fix the explicit form of
the regulator after we transfer into the
curvilinear coordinate system
in which the loop parameters $s,t$ play the
role of the coordinate. For example, we set
\beq
\utep_{ij}\dot{\alpha}^{i}(s)\dot{\beta}^{j}(t)
\tilde{f}_{\ep}(\alpha(s),\beta(t))=
\frac{\sigma(\alpha,\beta)}{4\ep^{2}}
\theta(\ep-|s-s_{0}|)\theta(\ep-|t-t_{0}|),\label{eq:regulator}
\eeq
where $\sigma(\alpha,\beta)$ stands for
the signature which takes the value
$+1$($-1$) if $(\dot{\alpha}(s_{0}),\dot{\beta}(t_{0}))$ forms a
right-(left-)handed frame. Putting these ingredients
into together and taking the limit
$\ep\rightarrow0$, we finally find:
\begin{eqnarray}
\hat{\cal H}_{1}(\tN)
(h_{\alpha}[0,s_{0}]h_{\beta}[t_{0},1])_{A}\UI{B}
&=&-\frac{1}{8}\sigma(\alpha,\beta)\tN(x_{0})
(h_{\alpha}[0,s_{0}]\tB(x_{0})h_{\beta}[t_{0},1])_{A}\UI{B}
\label{eq:kink}\\
&=&-\frac{1}{8}\sigma(\alpha,\beta)\tN(x_{0})
\AD(\alpha_{1}\circ\beta_{2},x_{0})
(h_{\alpha}[0,s_{0}]h_{\beta}[t_{0},1])_{A}\UI{B},\nonumber
\end{eqnarray}
where $\AD$ is the area derivative
which acts on the functionals of graphs
\footnote{We regard $\alpha$ as a
part of a graph. $\gamma_{x_{0}}$ is
an infinitesimal loop with a basepoint $x_{0}$.
The \lq coordinate area'
$\UT{s}(\gamma)$ of a loop $\gamma$ is defined
by: $\UT{s}(\gamma)\equiv\frac{1}{2}
\utep_{ij}\oint\gamma^{i}d\gamma^{j}$.}:
$$
\AD(\gamma,x_{0})\Psi[\alpha,\cdots]\equiv
\lim_{\UT{s}\rightarrow0}
\frac{\Psi[\alpha\circ\gamma_{x_{0}},\cdots]
-\Psi[\alpha,\cdots]}{\UT{s}(\gamma_{x_{0}})}.
$$
To derive the last equality of eq.(\ref{eq:kink}),
we have used the result in
ref.\cite{migdal}.

The third example is a pair of smooth loops
intersecting at a vertex. The first
example shows that the action of $\hat{\cal H}_{1}(\tN)$ is zero.
Thus we have only to compute the action of $\hat{\cal H}_{2}(\tN)$:
\begin{eqnarray*}
& &\hat{\cal H}^{\ep}_{2}(\tN)\left(h_{\alpha}[0,1]_{A}\UI{B}
h_{\beta}[0,1]_{C}\UI{D}\right)\\
&=&\int_{0}^{1}ds\int_{0}^{1}dt
\utep_{ij}\dot{\alpha}^{i}\dot{\beta}^{j}
\tilde{f}_{\ep}(\alpha(s),\beta(t))\tN(x_{0})
{\rm Tr}(\tB(x_{0})\lambda^{b}\lambda^{c})\\
& &\times\left\{
(h_{\alpha}[0,s]\lambda_{b}h_{\alpha}[s,1])_{A}\UI{B}
(h_{\beta}[0,t]\lambda_{c}h_{\beta}[t,1])_{C}\UI{D}
-(b \leftrightarrow c)+O(\ep)\right\}.
\end{eqnarray*}
We can reduce the integrand by using the Fiertz identity
(\ref{eq:fiertz})
and eq.(\ref{eq:regulator}).
The final result is:
\begin{eqnarray}
&&\hat{\cal H}_{2}(\tN)\left(h_{\alpha}[0,1]_{A}\UI{B}
h_{\beta}[0,1]_{C}\UI{D}\right) \nonumber \\*
&=&\frac{\sigma(\alpha,\beta)}{4}\tN(x_{0})
\left\{(h_{\alpha}[0,s_{0}]h_{\beta}[t_{0},1])_{A}\UI{D}
(h_{\beta}[0,t_{0}]\tB(x_{0})h_{\alpha}[s_{0},1])_{C}\UI{B}\right.
\nonumber \\*
& &
\left.\qquad
-(h_{\alpha}[0,s_{0}]\tB(x_{0})h_{\beta}[t_{0},1])_{A}\UI{D}
(h_{\beta}[0,t_{0}]h_{\alpha}[s_{0},1])_{C}\UI{B}\right\}
\nonumber \\*
&=&\frac{\sigma(\alpha,\beta)}{4}\tN_{x_{0}}
\left(\AD(\beta_{1}\circ\alpha_{2},x_{0})
-\AD(\alpha_{1}\circ\beta_{2},x_{0})\right) \nonumber \\*
& &\quad
\times\left\{(h_{\alpha}[0,s_{0}]h_{\beta}[t_{0},1])_{A}\UI{D}
(h_{\beta}[0,t_{0}]h_{\alpha}[s_{0},1])_{C}\UI{B}\right\}.
\end{eqnarray}

The action on the rest of the configurations
can be calculated similarly.
We list the action of $\hat{\cal H}(\tN)$ on all the basic
configurations in Appendix. There we make
use of the graphical representation.

Since we have at hand the action of the Hamiltonian constraint on the
basic configurations, it is not difficult to calculate its action
on general spin network states. The idea is the following.
Appendix tells us that the action of $\hat{\cal H}(\tN)$ has
nonvanishing contributions only at the vertices.
Because the action of $\hat{\cal H}(\tN)$ is local,

we can separate its action on the individual vertices, i.e.
\beq
\hat{\cal H}(\tN)\Psi^{\rm spinnet}(\omega)=\sum_{v\in V}
\left(\hat{\cal H}(\tN)\Psi^{\rm spinnet}(\omega)\right)|_{v},
\label{eq:individual}
\eeq
where $V$ denotes the set of all vertices
involved in the graph on which
the spin network state $\Psi^{spinnet}$ is defined.

If we use the separation (\ref{eq:sepa}),
we can easily evaluate the action
on each vertex $v$.
First we add up the contributions from all the kinks
and thus obtain the action of
$\hat{\cal H}_{1}(\tN)$ on the vertex $v$.
We then compute the sum of the action
of $\hat{\cal H}_{2}(\tN)$ on all
the pairs of parallel propagators.
The total sum of these contributions yields
the action of $\hat{\cal H}(\tN)$ at the vertex $v$.
Symbolically we can write:
\begin{eqnarray}
\left(\hat{\cal H}(\tN)\Psi^{\rm spinnet}(\omega)\right)|_{v}&=&
\sum_{k\in K_{v}}\left(\begin{array}{l}
\mbox{the action of $\hat{\cal H}_{1}(\tN)$}\\
\mbox{ on a kink $k$}\end{array}\right)\nonumber \\*
& &\qquad+\sum_{p\in P_{v}}\left(\begin{array}{l}
\mbox{the action of $\hat{\cal H}_{2}(\tN)$}\\
\mbox{on a pair $p$ of propagators}\end{array}\right),
\label{eq:decomposition}
\end{eqnarray}
where $K_{v}$ ($P_{v}$) is the set of all the kinks (all the pairs
of parallel propagators) at the vertex $v$.
Using eqs.(\ref{eq:individual}),
(\ref{eq:decomposition}) and the equations
in Appendix, we can therefore
reduce the problem of evaluating the
action of the Hamiltonian constraint
on the spin network states to that of combinatorics.
This applies to more general operators which
involve the product of a finite number of momenta $\tE^{i}_{a}$.
Thus, by virtue of this property of the spin network,
the problems of gauge theories or those of quantum gravity
may be formulated in a graphical and
combinatorial manner which is somewhat
similar to that of a perturbation theory using Feynmann diagrams.

%%%%%%%%%%%%%%%%%%%%%%%%%%%%%%%%%%%%%%%%%%%%%%%%%%%%%%%%%%%%%%%%%

\subsection{Covariance under the diffeomorphisms}

In the last subsection we have obtained the
action of the regularized
Hamiltonian constraint on the spin network states.
The expression
does not explicitly depend on the background structure used to
define the regulator (\ref{eq:regulator}),
and so it is expected to be
covariant under the spatial diffeomorphisms.
We will demonstrate explicitly that this is indeed the case.

As an illustration we consider the action on a kink:
$$
\hat{U}(\phi)\hat{\cal H}(\tN)\hat{U}(\phi^{-1})\cdot
(h_{\alpha}[0,s_{0}]h_{\beta}[t_{0},1])_{A}\UI{B}.
$$
Using eq.(\ref{eq:DIFF}) and the following equation:$
\hat{U}(\phi)\tB(x)=
\det(\partial_{i}\phi^{-1}(x)^{j})\tB(\phi^{-1}(x))$
we can easily compute the above expression:
\begin{eqnarray*}
&\!\!\!{}\!\!\!&\hat{U}(\phi)
\hat{\cal H}(\tN)\hat{U}(\phi^{-1})\cdot
(h_{\alpha}[0,s_{0}]h_{\beta}[t_{0},1])_{A}\UI{B}  \\
&\!\!\!=\!\!\!&-\frac{\sigma(\phi\cdot\alpha,\phi\cdot\beta)}{8}
\tN(\phi(x_{0}))\hat{U}(\phi)
\cdot(h_{\phi\cdot\alpha}[0,s_{0}]
\tB(\phi(x_{0}))h_{\phi\cdot\beta}[t_{0},1])_{A}\UI{B} \\
&\!\!\!=\!\!\!&-\frac{\sigma(\phi\cdot\alpha,\phi\cdot\beta)}{8}
\tN(\phi(x_{0}))
\left(h_{\alpha}[0,s_{0}]
\left(\det(\partial_{i}\phi^{-1}(x)^{j})|_{x=\phi(x_{0})}
\tB(x_{0})\right)h_{\beta}[t_{0},1]\right)_{A}\UI{B}\\
&\!\!\!=\!\!\!&-\frac{\sigma(\phi\cdot\alpha,\phi\cdot\beta)}{8}
(\phi_{\ast}\tN)(x_{0})(h_{\alpha}[0,s_{0}]
\tB(x_{0})h_{\beta}[t_{0},1])_{A}\UI{B},
\end{eqnarray*}
where $(\phi_{\ast}\tN)(x)\equiv
\det^{-1}(\partial_{i}\phi^{j}(x))\tN(\phi(x))$.
This is the desired result.
Though we do not demonstrate explicitly,
we can derive this result also
using the expression involving the area derivative.

The proof is essentially the same also in the general case.
The action of $\hat{\cal H}(\tN)$ on
the spin network states can be always
expressed by the composite operation of
:i)the topological manipulation in which the

curves through the vertices are cut and rejoined with appropriate
relative weights; and ii) the action of
the area derivative multiplied by the value
of \lq lapse' $\tN(x_{v})$ at each vertex.
It is obvious that,
when conjugated by a diffeomorphism,
operation i) does not pick anything
while the change in operation ii) amounts to replacing $\tN$ by
$\phi_{\ast}\tN$. Thus we have proved the diffeomorphism covariance
of our regularized Hamiltonian constraint:
\beq
\hat{U}(\phi)\hat{\cal H}(\tN)\hat{U}(\phi^{-1})
=\hat{\cal H}(\phi_{\ast}\tN). \label{eq:covarinace}
\eeq
Finally we should mention that, if
we \lq differentiate' this equation,
we obtain the quantum commutator
version of the Poisson algebra (\ref{eq:CPA})
between the Hamiltonian
constraint and the diffeomorphism constraint.

%%%%%%%%%%%%%%%%%%%%%%%%%%%%%%%%%%%%%%%%%%%%%%%%%%%%%%%%%%%%%%%%%%%%%

\section{Solutions to the Hamiltonian constraint}

\ \ \ In this section we provide solutions
to the Hamiltonian constraint.
After briefly reviewing the solutions
which have been found so far, we
construct the \lq combinatorial solutions'
which are the generalization
of the solution found by Jacobson and
Smolin to the spin network states.

%%%%%%%%%%%%%%%%%%%%%%%%%%%%%%%%%%%%%%%%%%%%%%%%%%%%%%%%%%%%%%

\subsection{Topological solutions}

As we have seen in the last section,
the nonvanishing action of the Hamiltonian
constraint $\hat{\cal H}(\tN)$ on
spin network states necessarily involves the
area derivative $\AD(\alpha,x_{v})$ at the vertex $v$.
Hence the action of $\hat{\cal H}(\tN)$
vanishes if the action of the area
derivative is zero everywhere,
which means that the spin connection is flat:
\beq
\tB(x)^{a}=\frac{1}{2}\otep^{ij}F^{a}_{ij}(x)=0.
\eeq
We therefore find the following \lq distributional' solution:
\beq
\Psi^{\rm top.}(\omega)\equiv
\psi(\omega)\prod_{a,x}\delta(\tB^{a}(x)),
\label{eq:topological}
\eeq
where $\psi(\omega)$ is an arbitrary
(square-integrable) gauge-invariant
function of $\omega_{i}^{a}$.
This is the very solution of the constraints
in Witten's formalism of (2+1)-gravity\cite{witt}
with a vanishing cosmological constant:
$$
\hat{\Psi}^{a}(x)\Psi^{\rm top.}(\omega)
=\hat{\cal G}^{a}(x)\Psi^{\rm top.}(x)=0.
$$
Because this kind of solution has a
support only on the flat connections,
$\psi(\omega)$ amounts to the function on the moduli space of flat
$SL(2,{\bf R})$ connections on $\Sigma$ modulo gauge transformations.
In terms of spin network states, this solution depends only on the
homotopy classes of the \lq colored graphs'.
As a corollary, it follows that
this solution is invariant under the diffeomorphisms and so
this topological solution is the solution to all the constraint.
We can thus consider the Hilbert space of Witten's formalism as a
subspace of the Hilbert space of
(2+1)-dimensional Ashtekar formalism.

%%%%%%%%%%%%%%%%%%%%%%%%%%%%%%%%%%%%%%%%%%%%%%%%%%%%%%%%%%%%%%%%%%

\subsection{Trivial solutions}

One of the important results obtained in the last section is that
the nonvanishing contributions to the action of $\hat{\cal H}(\tN)$
are only from the vertices, i.e. the points where the analyticity
of the curves breaks down.
{}From this result we realize that, if we consider
the spin network states consisting only
of the Wilson loops along the
smooth loops $\{\alpha_{i}\}$ ($i=1,\cdots,I$)
without any intersection:
\beq
\Psi_{\{(\alpha_{i},\rho_{i})\}}(\omega)=
\prod_{i=1}^{I}W(\alpha_{i},\rho_{i}),\label{eq:trivial}
\eeq
then these states solve the Hamiltonian constraint. These states
are related with the solutions to the Hamiltonian constraint which
have been found in the loop representation\cite{smol}.

%%%%%%%%%%%%%%%%%%%%%%%%%%%%%%%%%%%%%%%%%%%%%%%%%%%%%%%%%%%%%%%%%%%

\subsection{Combinatorial solutions}

If we look at Appendix carefully,
we find that the linear combination:
\beq
h_{\alpha}[0,1]_{A}\UI{B}h_{\beta}[0,1]_{C}\UI{D}-2
(h_{\alpha}[0,s_{0}]h_{\beta}[t_{0},1])_{A}\UI{D}
(h_{\beta}[0,t_{0}]h_{\alpha}[s_{0},1])_{C}\UI{B}\label{eq:jacob}
\eeq
has a vanishing action of $\hat{\cal H}(\tN)$.
This is the solution given by
Jacobson and Smolin\cite{jacob}.
We therefore expect that, even if the graph has
vertices, some appropriate linear
combinations of the spin network states
defined on a graph may solve the Hamiltonian constraint.
Here we will find such \lq combinatorial solutions'
which are considered to be
generical within the configuration
of the graphs considered in this paper.

We have seen in eq.(\ref{eq:individual})
that $\hat{\cal H}(\tN)$ acts on each vertices independently.
In order to be a solution to the Hamiltonian constraint equation,
the spin network state must solve this constraint equation
{\em at every vertex.} We can construct a solution
by looking for  the intertwining operators
which give the vanishing action of $\hat{\cal H}(\tN)$ {\em at
each vertex}, and by gluing these solutions at adjacent
(but separate) vertices using an
adequate parallel propagator as a glue.
Thus we have only to concentrate on one vertex, say, at $x_{0}$.
As in Appendix we introduce the \lq rescaled Hamiltonian':
\beq
\hat{\cal H}^{\prime}_{x_{0}}
\equiv-\sigma(\alpha,\beta)\frac{8}{\tN(x_{0})}
\left(\mbox{ the action of $\hat{\cal H}(\tN)$ at $x_{0}$}\right).
\eeq

In searching for the combinatorial solutions, the following identity
proves to be useful:

%%%%%%%%%%%%(identity(I))%%%%%%%%%%%%%%%%%%%%%%%%%%%%%%%%%%%%%%%%%
\begin{figure}[h]
\begin{picture}(150,70)
\put(0,40){\usebox{\LBRA}}
\put(5,55){\line(1,0){8}}
\put(17,55){\vector(1,0){8}}
\put(13,45){\line(0,1){8}}
\put(13,55){\vector(0,1){10}}
\put(17,45){\line(0,1){10}}
\put(17,57){\vector(0,1){8}}
\put(13,53){\line(1,1){4}}
\put(17,55){\circle*{1.5}}
{\scriptsize
\put(5,55){$A$}
\put(25,55){$B$}
\put(13,42){$C$}
\put(13,65){$D$}
\put(17,42){$E$}
\put(17,65){$F$}}
\put(25,40){\makebox(10,30){$-$}}
\put(35,55){\line(1,0){8}}
\put(47,55){\vector(1,0){8}}
\put(43,45){\line(0,1){8}}
\put(43,55){\vector(0,1){10}}
\put(47,45){\line(0,1){10}}
\put(47,57){\vector(0,1){8}}
\put(43,53){\line(1,1){4}}
\put(43,55){\circle*{1.5}}
{\scriptsize
\put(35,55){$A$}
\put(55,55){$B$}
\put(43,42){$C$}
\put(43,65){$D$}
\put(47,42){$E$}
\put(47,65){$F$}}
\put(55,40){\makebox(10,30){$+$}}
\put(65,55){\line(1,0){8}}
\put(77,55){\vector(1,0){8}}
\put(73,45){\line(0,1){6}}
\put(73,57){\vector(0,1){8}}
\put(77,45){\line(0,1){8}}
\put(77,59){\vector(0,1){6}}
\put(73,55){\line(1,1){4}}
\put(73,51){\line(1,1){4}}
\put(73,57){\line(1,-1){4}}
\put(77,55){\circle*{1.5}}
{\scriptsize
\put(65,55){$A$}
\put(85,55){$B$}
\put(73,42){$C$}
\put(73,65){$D$}
\put(77,42){$E$}
\put(77,65){$F$}}
\put(85,40){\makebox(10,30){$-$}}
\put(95,55){\line(1,0){8}}
\put(107,55){\vector(1,0){8}}
\put(103,45){\line(0,1){6}}
\put(103,57){\vector(0,1){8}}
\put(107,45){\line(0,1){8}}
\put(107,59){\vector(0,1){6}}
\put(103,55){\line(1,1){4}}
\put(103,51){\line(1,1){4}}
\put(103,57){\line(1,-1){4}}
\put(103,55){\circle*{1.5}}
{\scriptsize
\put(95,55){$A$}
\put(115,55){$B$}
\put(103,42){$C$}
\put(103,65){$D$}
\put(107,42){$E$}
\put(107,65){$F$}}
\put(115,40){\usebox{\RBRA}}

\put(0,0){\makebox(10,30){$=$}}
\put(10,0){\usebox{\LBRA}}
\put(15,16){\line(1,0){7}}
\put(24,15){\vector(1,0){11}}
\put(24,5){\line(0,1){10}}
\put(27,5){\vector(0,1){20}}
\put(22,16){\vector(0,1){9}}
\put(24,15){\circle*{1.5}}
{\scriptsize
\put(15,16){$A$}
\put(35,15){$B$}
\put(24,2){$C$}
\put(22,25){$D$}
\put(27,2){$E$}
\put(27,25){$F$}}
\put(35,0){\makebox(10,30){$-$}}
\put(45,16){\line(1,0){7}}
\put(54,15){\vector(1,0){11}}
\put(54,5){\line(0,1){10}}
\put(57,5){\vector(0,1){20}}
\put(52,16){\vector(0,1){9}}
\put(52,16){\circle*{1.5}}
{\scriptsize
\put(45,16){$A$}
\put(65,15){$B$}
\put(54,2){$C$}
\put(52,25){$D$}
\put(57,2){$E$}
\put(57,25){$F$}}
\put(65,0){\makebox(10,30){$+$}}
\put(75,15){\line(1,0){11}}
\put(88,14){\vector(1,0){7}}
\put(88,5){\line(0,1){9}}
\put(83,5){\vector(0,1){20}}
\put(86,15){\vector(0,1){10}}
\put(88,14){\circle*{1.5}}
{\scriptsize
\put(75,15){$A$}
\put(95,14){$B$}
\put(83,2){$C$}
\put(83,25){$D$}
\put(88,2){$E$}
\put(86,25){$F$}}

\put(95,0){\makebox(10,30){$-$}}
\put(105,15){\line(1,0){11}}
\put(118,14){\vector(1,0){7}}
\put(118,5){\line(0,1){9}}
\put(113,5){\vector(0,1){20}}
\put(116,15){\vector(0,1){10}}
\put(116,15){\circle*{1.5}}
{\scriptsize
\put(105,15){$A$}
\put(125,14){$B$}
\put(113,2){$C$}
\put(113,25){$D$}
\put(118,2){$E$}
\put(116,25){$F$}}
\put(125,0){\usebox{\RBRA}}
\put(130,0){\makebox(15,30){. (I)}}
\end{picture}
\end{figure}
%%%%%%%%%%%%%%%%%%%%%%%%%%%%%%%%%%%%%%%%%%%%%%%%%

This identity is proved as follows.
We can write down the difference between
the l.h.s and the r.h.s of identity (I) as:
\begin{eqnarray*}
\mbox{(l.h.s) $-$ (r.h.s)}&=&
h_{\beta}[0,t_{0}]_{A}\UI{G}h_{\beta}[t_{0},1]_{H}\UI{B}
h_{\alpha}[0,s_{0}]_{C}\UI{I}h_{\alpha}[s_{0},1]_{J}\UI{D}
h_{\alpha}[0,s_{0}]_{E}\UI{K}h_{\alpha}[s_{0},1]_{L}\UI{F}\\
& &\times \tB_{GIK}^{HJL},\\
\tB_{GIK}^{HJL}&\equiv&
\tB_{K}\UI{H}(\ep_{G}\UI{J}\ep_{I}\UI{L}
-\ep_{G}\UI{L}\ep_{I}\UI{J})
-\tB_{G}\UI{J}(\ep_{I}\UI{L}\ep_{K}\UI{H}
-\ep_{I}\UI{H}\ep_{K}\UI{L})\\
& &+\tB_{I}\UI{H}(\ep_{K}\UI{J}\ep_{G}\UI{L}
-\ep_{K}\UI{L}\ep_{G}\UI{J})
-\tB_{G}\UI{L}(\ep_{I}\UI{H}\ep_{K}\UI{J}
-\ep_{I}\UI{J}\ep_{K}\UI{H}).
\end{eqnarray*}
Using the two-spinor identity (\ref{eq:2spi}),
we can reduce $\tB_{GIK}^{HJL}$ to zero:
\begin{eqnarray*}
\tB_{GIK}^{HJL}&=&
\tB_{K}\UI{H}\ep_{GI}\ep^{JL}-\tB_{G}^{J}\ep_{IK}\ep^{LH}
+\tB_{I}\UI{H}\ep_{KG}\ep^{JL}-\tB_{G}\UI{L}\ep_{IK}\ep^{HJ}\\
&=&\ep_{IK}\tB_{G}^{H}\ep^{JL}-\ep^{JL}\tB_{G}\UI{H}\ep_{IK}=0.
\end{eqnarray*}
Thus we have shown that identity (I) holds.

Now we are in the place where we demonstrate explicitly
there exist a set of solutions to $\hat{\cal H}^{\prime}_{x_{0}}$
which are the extensions of the
solution found by Jacobson and Smolin to (four-valent) spin
network states.
These solutions are given by the following expression,
\beq
\prod_{i=1}^{m}(h_{\alpha}[0,s_{0}]_{A_{i}}\UI{E_{i}}
h_{\alpha}[s_{0},1]_{F_{i}}\UI{B_{i}})
\prod_{j=1}^{n}(h_{\beta}[0,t_{0}]_{C_{j}}\UI{G_{j}}
h_{\beta}[t_{0},1]_{H_{j}}\UI{D_{j}})\times
I^{\circ}(m,n)_{E_{1}\cdots E_{m};G_{1}\cdots G_{n}}
^{F_{1}\cdots F_{m};H_{1}\cdots H_{n}},\label{eq:combisol}
\eeq
where $I^{\circ}(m,n)$ is the relevant intertwining operator:
\begin{eqnarray}
I^{\circ}(m,n)_{E_{1}\cdots E_{m};G_{1}\cdots G_{n}}
^{F_{1}\cdots F_{m};H_{1}\cdots H_{n}}&\equiv&
\sum_{r=0}^{\min(m,n)}
\left(\prod_{i=1}^{r}\frac{-2}{m+n-2i+1}\right)
I^{r}(m,n)_{E_{1}\cdots E_{m};G_{1}\cdots G_{n}}
^{F_{1}\cdots F_{m};H_{1}\cdots H_{n}},
\nonumber \\*
I^{r}(m,n)_{E_{1}\cdots E_{m};G_{1}\cdots G_{n}}
^{F_{1}\cdots F_{m};H_{1}\cdots H_{n}}&\equiv&
\sum_{\scriptsize\begin{array}{l}
1\leq k_{1}<\cdots<k_{r}\leq m\\
1\leq l_{1}<\cdots<l_{r}\leq n
\end{array}}
\sum_{\sigma\in P_{r}}\nonumber \\*
&\times&\left(
\prod_{i=1}^{r}
\ep_{E_{k_{i}}}\UI{H_{l_{\sigma_{i}}}}
\ep_{G_{l_{\sigma_{i}}}}\UI{F_{k_{i}}}
\prod_{\scriptsize\begin{array}{c}
k^{\prime}=1\\ k^{\prime}\neq k_{1},\cdots,k_{r}
\end{array}}^{m}\ep_{E_{k^{\prime}}}\UI{F_{k^{\prime}}}
\prod_{\scriptsize\begin{array}{c}
l^{\prime}=1\\ l^{\prime}\neq l_{1},\cdots,l_{r}
\end{array}}^{n}\ep_{G_{l^{\prime}}}\UI{H_{l^{\prime}}}
\right).\label{eq:intertwiner}
\end{eqnarray}
$P_{r}$ in the above expression is the group of the permutations of
$r$ numbers $(l_{1},\cdots,l_{r})$.

{\it\large(Proof):} To simplify the calculation we first define the
following two sets of functionals:
\begin{eqnarray}
C^{r}(m,n)_{A_{1}\cdots A_{m};C_{1}\cdots C_{n}}
^{B_{1}\cdots B_{m};D_{1}\cdots D_{n}}&\!\!\!\equiv\!\!\!&
\prod_{i=1}^{m}(h_{\alpha}[0,s_{0}]_{A_{i}}\UI{E_{i}}
h_{\alpha}[s_{0},1]_{F_{i}}\UI{B_{i}})
\prod_{j=1}^{n}(h_{\beta}[0,t_{0}]_{C_{j}}\UI{G_{j}}
h_{\beta}[t_{0},1]_{H_{j}}\UI{D_{j}}) \nonumber \\*
&\!\!\!{}\!\!\!&\times
I^{r}(m,n)_{E_{1}\cdots E_{m};G_{1}\cdots G_{n}}
^{F_{1}\cdots F_{m};H_{1}\cdots H_{n}},
\label{eq:configBC} \\
B^{r}(m,n)_{A_{1}\cdots A_{m};C_{1}\cdots C_{n}}
^{B_{1}\cdots B_{m};D_{1}\cdots D_{n}}&\!\!\!\equiv\!\!\!&
\prod_{i=1}^{m}(h_{\alpha}[0,s_{0}]_{A_{i}}\UI{E_{i}}
h_{\alpha}[s_{0},1]_{F_{i}}\UI{B_{i}})
\prod_{j=1}^{n}(h_{\beta}[0,t_{0}]_{C_{j}}\UI{G_{j}}
h_{\beta}[t_{0},1]_{H_{j}}\UI{D_{j}})
\nonumber \\*
&\!\!\!\times\!\!\!&\sum_{\scriptsize\begin{array}{l}
1\leq k_{1}<\cdots<k_{r}\leq m\\
1\leq l_{1}<\cdots<l_{r}\leq n\end{array}}
\sum_{\sigma\in P_{r}}\sum_{i=1}^{r}
\prod_{\scriptsize\begin{array}{l}
j=1\\j\neq i\end{array}}^{r}
\ep_{E_{k_{i}}}\UI{H_{l_{\sigma_{i}}}}
\ep_{G_{l_{\sigma_{i}}}}\UI{F_{k_{i}}}
\nonumber \\*
&\!\!\!\times\!\!\!&(\tB_{E_{k_{i}}}\UI{H_{l_{\sigma_{i}}}}
\ep_{G_{l_{\sigma_{i}}}}\UI{F_{k_{i}}}
-\ep_{E_{k_{i}}}\UI{H_{l_{\sigma_{i}}}}
\tB_{G_{l_{\sigma_{i}}}}\UI{F_{k_{i}}})
\prod_{\scriptsize\begin{array}{c}
k^{\prime}=1\\ \mbox{\tiny $k^{\prime}\neq k_{1},\cdots,k_{r}$}
\end{array}}^{m}\ep_{E_{k^{\prime}}}\UI{F_{k^{\prime}}}
\prod_{\scriptsize\begin{array}{c}
l^{\prime}=1\\ \mbox{\tiny$l^{\prime}\neq l_{1},\cdots,l_{r}$}
\end{array}}^{n}\ep_{G_{l^{\prime}}}\UI{H_{l^{\prime}}}.\nonumber
\end{eqnarray}
We should notice that $B^{0}(m,n)=B^{\min(m,n)+1}(m,n)=0$.
Intuitively, the $C^{r}(n,m)$-functional is obtained as follows:
Prepare $m$ propagators $h_{\alpha}$ along $\alpha$ and $n$
propagators $h_{\beta}$ along $\beta$; then choose $r$ pairs of
$h_{\alpha}$ and $h_{\beta}$; cut and reglue
each pair in the orientation
preserving fashion; and finally sum up the
results obtained from all the
choices of $r$ pairs. The $B^{r}(m,n)$-functional is the result of
the action of $\Hp_{1}$ on the $C^{r}(m,n)$-functional.

We will henceforth omit the \lq external'
spinor indices each of which
is associated with the tip or the tail of a parallel propagator
$h_{\alpha}[0,1]$ or $h_{\beta}[0,1]$,
because they only serve as the labels of the propagators and  do
not play any essential role.
Graphical representations of $C^{r}(m,n)$ and $B^{r}(m,n)$ are
provided in figure 4.
There, we distinct the propagators by numbering them.

%%%%%%%%%%%%(figure 4)%%%%%%%%%%%%%%%%%%%%%%%%%%%%%%%%%%%%%%%%%
\begin{figure}[t]
\begin{center}
\begin{picture}(150,110)(-10,-20)
%%%%%def of C%%%%%%%%%%%%%%%%%
\put(0,50){\makebox(15,30){$C^{r}(m,n)$}}
\put(15,50){\makebox(10,30){$=$}}
\put(25,65){\LARGE$\sum$}
\put(20,60){\scriptsize $1\leq k_{1}<\cdots<k_{r}\leq m$}
\put(20,55){\scriptsize$1\leq l_{1}<\cdots<l_{r}\leq n$}
\put(55,65){\LARGE$\sum$}
\put(55,60){\scriptsize $\sigma\in P_{r}$}
\put(65,55){\vector(1,0){30}}
\put(65,75){\vector(1,0){30}}
\put(65,70){\line(1,0){10}}
\put(65,62){\line(1,0){18}}
\put(77,68){\vector(1,0){18}}
\put(85,60){\vector(1,0){10}}
\put(70,50){\vector(0,1){30}}
\put(90,50){\vector(0,1){30}}
\put(77,50){\line(0,1){18}}
\put(85,50){\line(0,1){10}}
\put(75,70){\vector(0,1){10}}
\put(83,62){\vector(0,1){18}}
{\footnotesize
\put(70,80){$1$}
\put(90,80){$m$}
\put(75,80){$k_{1}$}
\put(77,80){\makebox(5,5)[b]{$\cdots$}}
\put(83,80){$k_{r}$}
\put(95,75){$1$}
\put(95,55){$n$}
\put(95,68){$l_{\sigma_{1}}$}
\put(95,63){$\vdots$}
\put(95,60){$l_{\sigma_{r}}$}
}
%%%%%%%%%%%%%%%%%%%%%%%%%%
\put(100,50){\makebox(5,30){,}}

%%%%%%%def of B%%%%%%%%%%
\put(0,25){\makebox(15,10){$B^{r}(m,n)$}}
\put(15,25){\makebox(10,10){$=$}}
\put(25,28){\LARGE$\sum$}
\put(20,23){\scriptsize $1\leq k_{1}<\cdots<k_{r}\leq m$}
\put(20,18){\scriptsize$1\leq l_{1}<\cdots<l_{r}\leq n$}
\put(55,28){\LARGE$\sum$}
\put(55,23){\scriptsize $\sigma\in P_{r}$}
\put(68,28){\LARGE$\sum$}
\put(70,23){\scriptsize $i=1$}
\put(73,32){\scriptsize $r$}
\put(80,28){$\times$}
\put(25,-20){\makebox(10,30){$\times$}}
\put(35,-20){\usebox{\LBRA}}
%%%%%%%%%%%%%%%%%%%%%%
\put(45,-20){\vector(0,1){30}}
\put(65,-20){\vector(0,1){30}}
\put(51,-20){\line(0,1){19}}
\put(56,-20){\line(0,1){14}}
\put(61,-20){\line(0,1){9}}
\put(49,1){\vector(0,1){9}}
\put(54,-4){\vector(0,1){14}}
\put(59,-9){\vector(0,1){19}}
\put(40,5){\vector(1,0){30}}
\put(40,-15){\vector(1,0){30}}
\put(40,1){\line(1,0){9}}
\put(40,-4){\line(1,0){14}}
\put(40,-9){\line(1,0){19}}
\put(51,-1){\vector(1,0){19}}
\put(56,-6){\vector(1,0){14}}
\put(61,-11){\vector(1,0){9}}
\put(56,-6){\circle*{1.5}}
{\scriptsize
\put(45,10){$1$}
\put(65,10){$m$}
\put(49,10){$k_{1}$}
\put(51.5,10){$\cdot\cdot$}
\put(54,10){$k_{i}$}
\put(56.5,10){$\cdot\cdot$}
\put(59,10){$k_{r}$}
\put(70,5){$1$}
\put(70,-15){$n$}
\put(70,-1){$l_{\sigma_{1}}$}
\put(70,-3){$\vdots$}
\put(70,-6){$l_{\sigma_{i}}$}
\put(70,-8){$\vdots$}
\put(70,-11){$l_{\sigma_{r}}$}
}
%%%%%%%%%%%%%%%%%%%%%%
\put(75,-20){\makebox(10,30){$-$}}

\put(90,-20){\vector(0,1){30}}
\put(110,-20){\vector(0,1){30}}
\put(96,-20){\line(0,1){19}}
\put(101,-20){\line(0,1){14}}
\put(106,-20){\line(0,1){9}}
\put(94,1){\vector(0,1){9}}
\put(99,-4){\vector(0,1){14}}
\put(104,-9){\vector(0,1){19}}
\put(85,5){\vector(1,0){30}}
\put(85,-15){\vector(1,0){30}}
\put(85,1){\line(1,0){9}}
\put(85,-4){\line(1,0){14}}
\put(85,-9){\line(1,0){19}}
\put(96,-1){\vector(1,0){19}}
\put(101,-6){\vector(1,0){14}}
\put(106,-11){\vector(1,0){9}}
\put(99,-4){\circle*{1.5}}
{\scriptsize
\put(90,10){$1$}
\put(110,10){$m$}
\put(94,10){$k_{1}$}
\put(96.5,10){$\cdot\cdot$}
\put(99,10){$k_{i}$}
\put(101.5,10){$\cdot\cdot$}
\put(104,10){$k_{r}$}
\put(115,5){$1$}
\put(115,-15){$n$}
\put(115,-1){$l_{\sigma_{1}}$}
\put(115,-3){$\vdots$}
\put(115,-6){$l_{\sigma_{i}}$}
\put(115,-8){$\vdots$}
\put(115,-11){$l_{\sigma_{r}}$}
}

\put(120,-20){\usebox{\RBRA}}
\put(125,-20){\makebox(5,30){.}}

\end{picture}
\end{center}
\caption{Graphical representations
of $C^{r}(m,n)$ and $B^{r}(m,n)$}
\end{figure}
%%%%%%%%%%%%%%%%%%%%%%%%%%%%%%%%%%%%%%%%%%%%%%%%%

Now the action of $\hat{\cal H}^{\prime}_{x_{0}}$ on $C^{r}(m,n)$ is
computed as:
\beq
\hat{\cal H}^{\prime}_{x_{0}}C^{r}(m,n)
=(m+n-2r+1)B^{r}(m,n)+2B^{r+1}(m,n).
\label{eq:identity3}
\eeq
We will only give a sketch of the proof of the above equation.
The contributions to the action of $\hat{\cal H}^{\prime}_{x_{0}}$ on
$C^{r}(m,n)$ is classified into four types (see Appendix):
i) contributions of type (A.2);
ii)contributions of type (A.3);
iii) contributions of type (A.4) and those
of type (A.5); and iv) is the same as iii) except that
$\alpha$ being replaced by $\beta$.

We can easily show that the class i)-contributions and the class
ii)-contributions sum up to yield $B^{r}(m,n)$ and $2B^{r+1}(m,n)$
respectively.

It is somewhat difficult to evaluate class iii)- and class
iv)-contributions. First we see that,
in the sum over the contributions
belonging to each of the two classes, the terms like the second term
in the r.h.s of eq.(A.4) and those like the first term in the r.h.s
of eq.(A.5) cancel out. Then we can apply identity (I) to the
surviving contributions, the sum of these terms turns out to be
proportional to $B^{r}(m,n)$. A careful consideration on permutations
and combinations tells us that the class iii)- and the class iv)-
contributions respectively add up to yield $(m-r)B^{r}(m,n)$ and
$(n-r)B^{r}(m,n)$ respectively. Thus the total sum of all the
contributions to $\hat{\cal H}^{\prime}_{x_{0}}C^{r}(m,n)$ is:
$$
B^{r}(m,n)+2B^{r+1}(m,n)+(m-r+n-r)B^{r}(m,n)
=(m+n-2r+1)B^{r}(m,n)+2B^{r+1}(m,n).
$$
This is nothing but the r.h.s of (\ref{eq:identity3}).

Now we have derived eq.(\ref{eq:identity3}),
it is an elementary exercise of
the linear algebra to show the following equation holds:
\beq
\hat{\cal H}^{\prime}_{x_{0}}\left(
\sum_{r=0}^{\min(m,n)}\left(\prod_{i=1}^{r}\frac{-2}{m+n-2i+1}\right)
C^{r}(m,n)\right)=0.
\eeq
Separating the expression in the
large parenthesis into the product of
parallel propagators along $\alpha_{1}$, $\beta_{1}$, $\alpha_{2}$
and $\beta_{2}$ and the intertwining operator, we obtain
eq.(\ref{eq:combisol}).  \hfill $\Box$

Let us now consider whether there are any combinatorial solutions
other than eq.(\ref{eq:combisol}). The key point in finding these
combinatorial solutions is whether there exist more than two
configurations, or graphs, such that the action of
$\hat{\cal H}^{\prime}_{x_{0}}$
on them cancel with each other.
At the four point vertex, we have such
configurations like the two terms in eq.(\ref{eq:jacob}).
It is obvious from eqs.(A.2)(A.3) we cannot find
any combinatorial solutions from two point vertices.

Similarly at three point vertices,
no nontrivial configuration seems to
exist on which the total action of
$\hat{\cal H}^{\prime}_{x_{0}}$ vanishes.
Hence we conjecture that\\
{\it(Conjecture1): We cannot construct any nontrivial
combinatorial solutions

which involve two or three point vertices.}

Thus it is sufficient to consider only four point vertices.
If we closely look at the proof of
eq.(\ref{eq:combisol}), it seems to be
essential that, in a configuration,
the parallel propagators along the curves
$\alpha_{1}\circ\beta_{2}$ and those along
$\beta_{1}\circ\alpha_{1}$ appear
in pairs. This indicates that,
in a configuration which is a part of
a combinatorial solution,
the number of the parallel propagators
along each smooth curve is conserved at each vertex. \\
{\it(Conjecture2): In any nontrivial combinatorial solutions,
the configurations at each vertex are
constructed by cutting and rejoining
the parallel propagators along the smooth
curves through that vertex. }

If we allow only configurations
appearing in conjecture2 , it is almost
evident that eq.(\ref{eq:combisol})
provides the combinatorial
solutions at a vertex $x_{0}$
which are general under the assumption
made in \S 2. The reasoning is as follows.
Let us consider to cancel the action of
$\hat{\cal H}^{\prime}_{x_{0}}$
on a configuration which appear in the sum in $C^{r}(m,n)$.
To cancel the contributions like
the second term in eq.(A.4) and the first
term in eq.(A.5), we have to sum up all the configurations in
$C^{r}(m,n)$. The problem is thus reduced to that of finding linear
combinations of $C^{r}(m,n)$ $(r=1,\cdots,\min(m,n))$ which
yield zero eigenvalue of $\hat{\cal H}^{\prime}_{x_{0}}$. We can
find immediately the unique solution, which is identical to
(\ref{eq:combisol}). Thus we conclude\\
{\it{\bf(Conjecture)}:
Eq.(\ref{eq:combisol}) yields the combinatorial
solutions which are general on the restricted graphs in which
at most two independent curves may intersect at each vertices.}

To prove these conjectures rigorously, of course, we need to
ensure that any \lq miraculous' cancellations can never occur
in general except those in eq.(\ref{eq:combisol}).
If these conjectures turn out to be true, we can extend them to
more general cases in which any types of vertices (except those
shown in figure 2(c), 2(d)) are considered.
Because the essence is in the cancellation
between the actions of $\hat{\cal H}^{\prime}_{x_{0}}$
on the two terms in eq.(\ref{eq:jacob}), we conjecture that
the combinatorial solutions consist only of the configurations
in which the number of the propagators
along each smooth curve is conserved
at every vertex. A corollary of this generalized conjecture
is that we cannot construct
a combinatorial solution from the trivalent graphs.
Thus to consider
the space of the solutions to all the constraints,
we would be forced to
consider $n$-valent graphs with $n\geq 4$.
While we have not yet found the proof, we expect that these conjectures
are indeed true.

Finally we provide the procedure to construct a combinatorial solution
which is also gauge invariant:\\
1) Choose a set of smooth loops ${\alpha_{i}}:[0,1]\rightarrow\Sigma$
$(i=1,2,\cdots,N; \alpha_{i}(0)=\alpha_{i}(1))$ in
the 2 dimensional space $\Sigma$.
Each loop in this set can intersect with
itself or with the others, provided that more than two loops do not
intersect at a point.\\
2) Equip each loop $\alpha_{i}$ with
an $l_{i}$-th power of spinor
Wilson loops: $(h_{\alpha_{i}}[0,1]_{A}\UI{A})^{l_{i}}$.\\
3) At each point where two loops, e.g.
$\alpha_{i}$ and $\alpha_{j}$
intersect, cut the spinor Wilson loops and
rejoin them by using as a glue
the intertwining operator $I^{\circ}(l_{i},l_{j})$ in
eq.(\ref{eq:intertwiner}).

We should note that eq.(\ref{eq:combisol})
remains to be the solution even if
we permutate the external indices.
Moreover, from eq.(\ref{eq:2spi2}),
we see that the antisymmetrized tensor product
of two spinor propagators
yields the invariant spinor. As a result
the antisymmetrization reduces the number
of the spinor propagators by two. We can therefore replace
procedure 2) by:\\
2$)^{\prime}$ Equip each loop $\alpha_{i}$ with a Wilson loop in the
spin-$\frac{l_{i}}{2}$ representation, which is given by the
symmetrized trace of $l_{i}$ spinor propagators:
$$
h_{\alpha_{i}}[0,1]_{A_{1}}\UI{(A_{1}}\cdots
h_{\alpha_{i}}[0,1]_{A_{l_{i}}}\UI{A_{l_{i}})}.
$$
We will denote the combinatorial solution constructed by
procedures 1), 2$)^{\prime}$ and 3) by
$\Psi^{combi.}_{\{(\alpha_{i},l_{i})\}}(\omega)$.
If we fix a set of loops $\{\alpha_{i}\}$ $(i=1,\cdots,N)$,
we can construct $({\bf Z}_{+})^{N}$ combinatorial solutions
where ${\bf Z}_{+}$ is the set of non-negative integers.
The trivial solutions described in the last subsection
are contained in the combinatorial solutions
$\Psi^{combi.}_{\{(\alpha_{i},l_{i})\}}(\omega)$
as particular configurations
in which all the loops $\alpha_{i}$ do not intersect.

%%%%%%%%%%%%%%%%%%%%%%%%%%%%%%%%%%%%%%%%%%%%%%%%%%%%%%%%%%%

\subsection{Solutions to all the constraints}

In the previous subsections we have found the solutions to
both the Gauss law and the Hamiltonian constraints.
To construct the physical Hilbert space which is spanned by
physical wavefunctions, however,
the diffeomorphism constraint remains
to be solved. Because the diffeomorphism
constraint equation tells us
that the physical states are invariant under the diffeomorphisms
$$
\hat{U}(\phi)\cdot\Psi^{phys.}(\omega)=\Psi^{phys.}(\omega),
\quad \phi\in {\rm Diff}(\Sigma),
$$
a prescription to construct such states from the
combinatorial solutions is to average
the action of the diffeomorphism operators on these states:
\begin{eqnarray}
\Psi^{phys.}_{\{(\alpha_{i},l_{i})\}}(\omega)
&=&\{Vol({\rm Diff}(\Sigma))\}^{-1}
\int_{{\rm Diff}(\Sigma)}[{\cal D}\phi]
\hat{U}(\phi)\cdot\Psi^{combi.}_{\{(\alpha_{i},l_{i})\}}
\nonumber \\
&=&\{Vol({\rm Diff}(\Sigma))\}^{-1}
\int_{{\rm Diff}(\Sigma)}[{\cal D}\phi]
\Psi^{combi.}_{\{(\phi\cdot\alpha_{i},l_{i})\}}\quad,
\end{eqnarray}
where $[{\cal D}\phi]$ is an \lq invariant measure on the space
${\rm Diff}(\Sigma)$
of the diffeomorphisms on $\Sigma$.
This average is formal in the sense that
it is indefinite because it involves
the ratio of two divergent expressions,
namely, $Vol({\rm Diff}(\Sigma))$ and
$\int_{{\rm Diff}(\Sigma)}[{\cal D}\phi]$.
An attempt to make this
formal average be mathematically rigorous
can be seen in ref.\cite{lewa}.
A rough outline of the strategy used there is the following.
Because there
is an infinite \lq isotropy group', i.e.
the group of diffeomorphisms which
leave invariant the set $\{(\alpha_{i},l_{i})\}$ of colored loops,
we can replace the formal average over
the entire group ${\rm Diff}(\Sigma)$
of diffeomorphisms by
a discrete sum on the orbit $[\{(\alpha_{i},l_{i})\}]$ of
$\{(\alpha_{i},l_{i})\}$ under ${\rm Diff}(\Sigma)$:
\beq
\Psi^{phys.}_{\{(\alpha_{i},l_{i})\}}(\omega)=
\sum_{\{(\alpha^{\prime}_{i},l_{i}^{\prime})\}
\in[\{(\alpha_{i},l_{i})\}]}
\Psi^{combi.}_{\{(\alpha^{\prime}_{i},l_{i}^{\prime})\}}(\omega).
\eeq
This is well-defined as a distributional
wavefunction which belongs to
the dual $\Phi^{\prime}$ of the space $\Phi\equiv Cyl^{\infty}
(\overline{\cal A/G})$ of the smooth cylindrical functions\footnote{
A cylindrical function is the function which is the projective
limit of the functions on the space of connections in a \lq lattice
gauge theory' defined on the graphs embedded in $\Sigma$. For
details of the Cylindrical function
and the projective limit, see e.g.
ref.\cite{asle}}. Hence, from the
combinatorial solutions $\Psi^{combi.}
_{\{(\alpha_{i,l_{i}})\}}$, we can construct the
physical wavefunctions $\Psi^{phys.}_{[\{(\alpha_{i},l_{i})\}]}$
which depend only on the differential topology of the set
$\{(\alpha_{i},l_{i})\}$ of
smooth colored loops embedded in $\Sigma$.

%%%%%%%%%%%%%%%%%%%%%%%%%%%%%%%%%%%%%%%%%%%%%%%%%%%%%%%%%%%%%%%%%

\section{Physical properties of the combinatorial solutions}

We have constructed infinitely many solutions
$\Psi^{combi.}_{\{(\alpha_{i},l_{i})\}}(\omega)$ to
the Hamiltonian constraint for each fixed set
$\{\alpha_{i}\}$ of smooth loops.
Let us now try to give physical interpretations
to these solutions. For this purpose we will consider
the action of a few types of operators on these solutions.

First we consider the \lq normal vector operator'
\beq
\hat{n}^{a}(\eta_{a})\equiv
\int_{\Sigma}d^{2}x\eta_{a}(x)\hat{\tn^{a}}(x)
=\int_{\Sigma}d^{2}x\eta_{a}
\frac{1}{2}\ep^{abc}\utep_{ij}
\hat{\tE}^{i}_{b}(x)\hat{\tE}^{j}_{c}(x)
\eeq
which is not gauge invariant.
This operator needs to be regularized
because it involves two functional derivatives at a point.
We adopt the same regularization
as that of the Hamiltonian constraint:
\begin{eqnarray}
\hat{n}^{a}(\eta_{a})&=&
\lim_{\ep\rightarrow0}(\hat{n}^{a}(\eta_{a}))^{\ep},
\nonumber \\
(\hat{n}^{a}(\eta))^{\ep}&\equiv&-\frac{1}{2}
\int_{\Sigma}d^{2}x
\int_{\Sigma}d^{2}y\tilde{f}_{\ep}(x,y)\utep_{ij}
\nonumber \\
&\times&{\rm Tr}
(h_{yx}[0,1]\eta(x)\lambda^{b}h_{xy}[0,1]\lambda^{c})
\frac{\delta}{\delta\omega^{b}_{i}(x)}
\frac{\delta}{\delta\omega^{c}_{j}(y)},
\label{eq:Normal}
\end{eqnarray}
where we have set $\eta\equiv\eta^{a}\lambda_{a}$.
This is the same expression as that of the regularized Hamiltonian
constraint with $\eta(x)$ being substituted for $8\tN(x)\tB(x)$.
The action of $\hat{n}^{a}(\eta_{a})$ can therefore be calculated
in the same way as that of computing
the action of $\hat{\cal H}(\tN)$.
When the combinatorial solutions were derived,
the action of $\tB$, or
the area derivative $\AD$, did not play any essential roles.
Thus each combinatorial solution gives zero eigenvalue to the
operator which is obtained by replacing
$\tN(x)\tB^{a}(x)$ by a certain
functional $F^{a}(\omega,x)$ of the spin connection.
The normal vector operator is one such operator.
Classically, vanishing of the \lq normal vector' $\tn^{a}$ indicates
that the induced metric $h_{ij}$ on $\Sigma$ is degenerate.
Hence we can consider that the combinatorial solutions correspond
to the classical solutions with degenerate metric
which do not belong to the solutions in Witten's formalism.

In practice, the \lq normal vector' operator is not so useful
because it is not invariant under gauge transformations.
Thus we have to consider some operators which
carry the gauge invariant information on $\tn^{a}$.
A useful candidate is provided by the operators
one of which measures the area of some region $D$ in $\Sigma$.
Classical description of the area is given by:
$$
S(D)=\int_{D}d^{2}x\sqrt{h(x)}=\int_{D}d^{2}x(-\tn^{a}\tn_{a})^{1/2}.
$$
In order to define its quantum version
we will mimic the prescription
for defining the operator version of
the volume or area in (3+1)-dimensional
gravity\cite{rove}. Dividing the region $D$
into small regions $D_{I}$
($I=1,2,\cdots$) whose coordinate areas
are supposed to be of order
$\delta^{2}$, we define the area operator as follows:
\begin{eqnarray}
\hat{S}(D)&=&\lim_{\delta\rightarrow0}\sum_{I}\hat{S}_{I}
\nonumber \\
(\hat{S}_{I})^{2}&\equiv&-\hat{S}_{I}^{a}\hat{S}_{I}^{b}\eta_{ab}
\nonumber \\
\hat{S}_{I}^{a}&\equiv&\int_{D_{I}}d^{2}x\hat{\tn}^{a}(x).
\label{eq:area}
\end{eqnarray}
Because the last expression involves two
functional derivatives at a point,
we further have to regularize this expression:
\begin{eqnarray}
\hat{S}_{I}^{a}&=&\lim_{\ep\rightarrow0}(\hat{S}_{I}^{a})^{\ep},
\nonumber \\
(\hat{S}_{I}^{a})^{\ep}&\equiv&
-\frac{1}{2}\int_{D_{I}}d^{2}x\int_{\Sigma}d^{2}y
\tilde{f}_{\ep}(x,y)
\utep_{ij}{\rm Tr}(h_{yx}\lambda^{a}\lambda^{b}h_{xy}\lambda{c})
\frac{\delta}{\delta\omega_{i}^{b}(x)}
\frac{\delta}{\delta\omega_{j}^{c}(y)}.
\label{eq;regarea}
\end{eqnarray}
This is nothing but eq.(\ref{eq:Normal})
with the smearing function
$\eta_{a}$ being the characteristic function:
$$
\eta_{a^{\prime}}(x)=\left\{\begin{array}{lll}
\delta_{a^{\prime}}^{a}&{\rm if}& x\in D_{I}\\
0                     &{\rm if}& x\in\Sigma\setminus D_{I}.
\end{array}\right.
$$
As a consequence, the action of the single
$\hat{S}_{I}^{a}$-operator
on the combinatorial solutions vanishes
$$
\hat{S}_{I}^{a}\cdot\Psi^{combi.}_{\{(\alpha_{i},l_{i})\}}(\omega)=0.
$$
{}From this result we can say that the naive
\lq squared infinitesimal area
operator' which is defined by\footnote{
$\lim_{\ep\rightarrow0}\lim_{\ep^{\prime}\rightarrow0}$
denotes the limit
in which we first take $\ep^{\prime}$
to zero and then $\ep$ to zero.}
\beq
(\hat{S}_{I})^{2}_{naive}\equiv-\lim_{\ep\rightarrow0}
\lim_{\ep^{\prime}\rightarrow0}
(\hat{S}_{I}^{a})^{\ep}(\hat{S}_{Ia})^{\ep^{\prime}}
=-\lim_{\ep^{\prime}\rightarrow0}
\lim_{\ep\rightarrow0}
(\hat{S}_{I}^{a})^{\ep}(\hat{S}_{Ia})^{\ep^{\prime}}
\label{eq:naivearea}
\eeq
has zero eigenvalues on the combinatorial solutions
\beq
(\hat{S}_{I})^{2}_{naive}\cdot
\Psi^{combi.}_{\{(\alpha_{i},l_{i})\}}(\omega)=0.
\label{eq:areavanish}
\eeq

A more careful consideration shows, however,
that the result depends on
the detail of the limit $\ep,\ep^{\prime}\rightarrow0$.
A candidate for the regularized

squared infinitesimal area operator
which we feel more natural than eq.(\ref{eq:naivearea})
in the field theoretical viewpoint is:
\begin{eqnarray}
(\hat{S}_{I})^{2}_{natural}&\equiv&
-\lim_{\ep=\ep^{\prime}\rightarrow0}
(\hat{S}_{I}^{a})^{\ep}(\hat{S}_{Ia})^{\ep^{\prime}}
\nonumber \\
&=&\frac{1}{32}\int_{D_{I}}d^{2}x\int_{\Sigma}d^{2}x^{\prime}
\int_{D_{I}}d^{2}y\int_{\Sigma}d^{2}y^{\prime}
\tilde{f}_{\ep}(x,x^{\prime})\tilde{f}_{\ep}(y,y^{\prime})
\utep_{ij}\utep_{kl}\nonumber \\
& &\times(\eta^{ac}\eta^{bd}+O(\ep,\ep^{\prime}))
\frac{\delta}{\delta\omega_{i}^{a}(x)}
\frac{\delta}{\delta\omega_{j}^{b}(x^{\prime})}
\frac{\delta}{\delta\omega_{k}^{c}(y)}
\frac{\delta}{\delta\omega_{l}^{d}(y^{\prime})}.
\label{eq:naturalarea}
\end{eqnarray}
Owing to eq.(\ref{eq:areavanish}),
when we are interested in the action on
the combinatorial solutions,
we have only to consider the difference:
\begin{eqnarray}
\delta(\hat{S}_{I})^{2}&\equiv&
(\hat{S}_{I})^{2}_{natural}-(\hat{S}_{I})^{2}_{naive}
\nonumber \\
&=&\{\lim_{\ep=\ep^{\prime}\rightarrow0}-\frac{1}{2}
(\lim_{\ep\rightarrow0}\lim_{\ep^{\prime}\rightarrow0}
+\lim_{\ep^{\prime}\rightarrow0}\lim_{\ep\rightarrow0})\}
(\hat{S}_{I}^{a})^{\ep}(\hat{S}_{Ia})^{\ep^{\prime}}.
\label{eq:diffarea}
\end{eqnarray}
Let us now consider the action of this difference
$\delta(\hat{S}_{I})^{2}$
on the spin network states\footnote{
Of course we assume that the intersection point under consideration
resides in the region $D_{I}$.}.
In our regularization introduced in \S 3,
due to the antisymmetric factor $\utep_{ij}$, the nonvanishing
contributions arise only when two functional derivatives,
\lq hands', from $(\hat{S}_{I}^{a})^{\ep}$ act on the parallel
propagators along independent curves.
The same discussion holds also for
$(\hat{S}_{I}^{a})^{\ep^{\prime}}$.
Thus we see that not more than two
hands act on one parallel propagator
along a smooth curve (or segment).
When two hands act on a smooth curve (or segment), one hand
comes from $(\hat{S}_{I}^{a})^{\ep}$ and the other from
$(\hat{S}_{I}^{a})^{\ep^{\prime}}$.
Next we consider the dependence on the limit taken.
In this case we are interested only
in the action of a pair of hands,
one of which is from $(\hat{S}_{I}^{a})^{\ep}$ and the other from
$(\hat{S}_{I}^{a})^{\ep^{\prime}}$, on one propagator.
We immediately see that the action of such pair on the propagator
which traverses the intersection along
a smooth curve is independent of
the way of taking the limit. Therefore, the only case depending on
how to take the limit turns out
to be the case in which such a pair of
hands act on the parallel propagator
along a segment which starts from
(or terminates at) the intersection considered.
The relation among the
limitation procedures is:

%%%%%%%%%%%%(identityII)%%%%%%%%%%%%%%%%%%%%%%%%%%%%%%%%%%%%%%%%%
\begin{figure}[ht]
\begin{picture}(160,30)(0,0)
%%%%%def of C%%%%%%%%%%%%%%%%%
\put(0,0){\makebox(5,30){\Huge$\{$}}
\put(5,13){\Large$\lim$}
\put(5,10){\scriptsize$\ep=\ep^{\prime}\rightarrow0$}
\put(15,0){\makebox(10,30){$-$}}
\put(25,0){\makebox(7,30){$\frac{1}{2}${\Large$($}}}
\put(30,13){\Large$\lim$}
\put(30,10){\scriptsize $\ep\rightarrow0$}
\put(40,13){\Large$\lim$}
\put(40,10){\scriptsize$\ep^{\prime}\rightarrow0$}
\put(50,0){\makebox(10,30){$+$}}
\put(60,13){\Large$\lim$}
\put(60,10){\scriptsize $\ep^{\prime}\rightarrow0$}
\put(70,13){\Large$\lim$}
\put(70,10){\scriptsize$\ep\rightarrow0$}
\put(75,0){\makebox(5,30)[r]{\Large$)$}}
\put(80,0){\makebox(5,30){\Huge$\}$}}
\put(90,0){\usebox{\LBRA}}
\put(100,5){\line(0,1){7,5}}
\put(100,13.5){\circle{2}}
\put(100,16){\circle*{2}}
\put(100,17.5){\vector(1,0){10}}
\put(110,0){\makebox(10,30){$+$}}
\put(125,5){\line(0,1){7,5}}
\put(125,13.5){\circle*{2}}
\put(125,16){\circle{2}}
\put(125,17.5){\vector(1,0){10}}
\put(140,0){\usebox{\RBRA}}
\put(145,0){\makebox(10,30){$=0$ ,}}
\put(153,0){(II)}
\end{picture}
\end{figure}
%%%%%%%%%%%%%%%%%%%%%%%%%%%%%%%%%%%%%%%%%%%%%%%%%
$\quad$\\
where a black small circle and a white one represent a hand from
$\delta(\hat{S}_{I}^{a})^{\ep}$ and one from
$\delta(\hat{S}_{Ia})^{\ep^{\prime}}$ respectively.

Using this relation, we conclude that
the nonvanishing contributions to
$\delta(\hat{S}_{I})^{2}$ are obtained
from only three configurations
which are depicted in figure 5, where the action of
$\delta(\hat{S}_{I})^{2}$ on these configurations is also given.
We should notice that, as a consequence of the above argument,
the action of $\delta(\hat{S}_{I})^{2}$ turns out to be divided into
two parts:
\beq
\delta(\hat{S}_{I})^{2}=\delta(\hat{S}_{I})^{2}|_{1}
+\delta(\hat{S}_{I})^{2}|_{2},
\eeq
where $\delta(\hat{S}_{I})^{2}|_{1}$ and
$\delta(\hat{S}_{I})^{2}|_{2}$
denote respectively the action on the single propagators and that
on the pairs of propagators.

%%%%%%%%%%%%%%% figure 5 %%%%%%%%%%%%%%%%%%%%%%%%%%%%%%%%%%%
\begin{figure}[t]
\begin{center}
\begin{picture}(150,130)(-15,-20)

%%%%%%%%%%% 1st fig.%%%%%%%%%%%%%%%%%%%%%%%%%%%%%
\put(0,80){\makebox(15,30){$\delta(\hat{S}_{I})^{2}|_{1}$}}
\put(15,80){\usebox{\LBRA}}
%%%begin lhs input%%%
\put(25,85){\line(0,1){10}}
\put(25,95){\vector(1,0){10}}
%%%end lhs input%%%

\put(40,80){\usebox{\RBRA}}
\put(45,80){\makebox(10,30){$=$}}
\put(55,80){\makebox(15,30){$-3\cdot 2^{-11}$}}
\put(70,80){\usebox{\LBRA}}

%%%begin rhs input%%%%
\put(80,95){\vector(1,0){10}}
\put(80,85){\line(0,1){10}}

%%%end rhs input%%%
\put(95,80){\usebox{\RBRA}}
\put(100,80){\makebox(5,30){,}}

%%%%%%%%%%% 2nd fig.%%%%%%%%%%%%%%%%%%%%%%%%%%%%%
\put(0,40){\makebox(15,30){$\delta(\hat{S}_{I})^{2}|_{1}$}}
\put(15,40){\usebox{\LBRA}}
%%%begin lhs input%%%
\put(24,45){\line(0,1){11}}
\put(24,56){\vector(1,0){11}}
\put(26,45){\line(0,1){9}}
\put(26,54){\vector(1,0){9}}

%%%end lhs input%%%

\put(40,40){\usebox{\RBRA}}
\put(45,40){\makebox(10,30){$=$}}
\put(55,40){\makebox(10,30){$-2^{-9}$}}
\put(65,40){\usebox{\LBRA}}

%%%begin rhs input%%%%
\put(74,54){\vector(1,0){11}}
\put(76,45){\line(0,1){11}}
\put(76,56){\vector(1,0){9}}
\put(74,45){\line(0,1){9}}
\put(90,40){\makebox(10,30){$-\frac{1}{2}$}}
\put(104,45){\line(0,1){11}}
\put(104,56){\vector(1,0){11}}
\put(106,45){\line(0,1){9}}
\put(106,54){\vector(1,0){9}}
%%%end rhs input%%%
\put(120,40){\usebox{\RBRA}}
\put(125,40){\makebox(5,30){,}}

%%%%%%%%%%% 3rd fig.%%%%%%%%%%%%%%%%%%%%%%%%%%%%%
\put(0,0){\makebox(15,30){$\delta(\hat{S}_{I})^{2}|_{1}$}}
\put(15,0){\usebox{\LBRA}}
%%%begin lhs input%%%
\put(29,16){\vector(0,1){9}}
\put(31,5){\line(0,1){9}}
\put(20,16){\line(1,0){9}}
\put(31,14){\vector(1,0){9}}

%%%end lhs input%%%

\put(40,0){\usebox{\RBRA}}
\put(45,0){\makebox(10,30){$=$}}
\put(55,0){\makebox(10,30){$2^{-9}$}}
\put(65,0){\usebox{\LBRA}}

%%%begin rhs input%%%%
\put(70,15){\vector(1,0){20}}
\put(80,5){\vector(0,1){20}}
\put(90,0){\makebox(10,30){$-\frac{1}{2}$}}
\put(109,16){\vector(0,1){9}}
\put(111,5){\line(0,1){9}}
\put(100,16){\line(1,0){9}}
\put(111,14){\vector(1,0){9}}
%%%end rhs input%%%
\put(120,0){\usebox{\RBRA}}
\put(125,0){\makebox(5,30){.}}
%%%%%%%%%%%%%%%%%%%%%%
\end{picture}
\end{center}
\caption{The basic actions of
$\delta(\hat{S}_{I})^{2}$ on spin network states}
\end{figure}
%%%%%%%%%%%%%%%%%%%%%%%%%%%%%%%%%%%%%%%%%%%%%%%%%

Now we have the basic actions in figure 5, we can calculate the
action of $(\hat{S}_{I})^{2}_{natural}$ on the combinatorial
solutions. We will only give its action on the simplest
solution eq.(\ref{eq:jacob}):
\beq
(\hat{S}_{I})^{2}_{natural}\left(
h_{\alpha}\otimes h_{\beta}
-2h_{\alpha_{1}\circ\beta_{2}}\otimes
h_{\beta_{1}\circ\alpha_{2}}\right)
=-2^{-8}\left(h_{\alpha}\otimes h_{\beta}
-2h_{\alpha_{1}\circ\beta_{2}}\otimes
h_{\beta_{1}\circ\alpha_{2}}\right),
\eeq
where we have simplified the notation.
{}From this result we expect
that the combinatorial solutions $\Psi^{combi.}$ have nonzero
eigenvalues of $(\hat{S}_{I})^{2}_{natural}$. We can regard this
as a sort of quantum effect because the classical counterpart
of $\Psi^{combi.}$ has degenerate metric
and so its area factor vanishes.

We can also construct the operator
which measures the length $L(\eta)$ of
a curve $\eta$ embedded in $\Sigma$. The classical expression
of $L(\eta)$ is
\beq
L(\eta)=
\int_{\eta}ds\sqrt{h_{ij}\dot{\eta}^{i}(s)\dot{\eta}^{j}(s)},
\eeq
where $h_{ij}=\utep_{ik}\utep_{jl}\tE^{k}_{a}\tE^{l}_{b}\eta^{ab}$
is the induced metric on $\Sigma$.
We can define its quantum operator version in the way
which is identical to that of defining the \lq area' operators
in (3+1)-dimensional gravity\cite{rove}.
We divide the curve $\eta$ into small
segments $\eta_{I}$ with length of order
$\delta$ and define $\hat{L}(\eta)$ as
\begin{eqnarray}
\hat{L}(\eta_{I})&=&\lim_{\delta\rightarrow0}
\sum_{I}\hat{L}(\eta_{I})
\nonumber \\
(\hat{L}(\eta_{I}))^{2}&\equiv&-\frac{1}{2}
\int_{\eta_{I}}d\sigma\dot{\eta}_{I}^{i}(\sigma)\utep_{ik}
\int_{\eta_{I}}d\tau\dot{\eta}_{I}^{j}(\tau)\utep_{jl}
\nonumber \\
& &\times{\rm Tr}
(\lambda^{a}h_{\sigma\tau}[0,1]\lambda^{b}h_{\tau\sigma}[0,1])
\frac{\delta}{\delta\omega_{k}^{a}(\eta_{I}(\sigma))}
\frac{\delta}{\delta\omega_{l}^{b}(\eta_{I}(\tau))}.
\label{eq:length}
\end{eqnarray}
Because this operator involves two functional derivatives, we can
separate its action as:
\beq
(\hat{L}(\eta_{I}))^{2}=(\hat{L}(\eta_{I}))^{2}|_{1}
+(\hat{L}(\eta_{I}))^{2}|_{2},\label{eq:lengthsepa}
\eeq
where the first term in the r.h.s

is the action on the single propagators
and the second one denotes the action on the pairs of propagators.
In principle we can calculate the
action even when the curve $\eta$
passes through the intersections.
For simplicity, however, we will only consider the case
where the segment $\eta_{I}$ intersects
with a curve $\alpha$ at the point
which is away from the intersections.
The basic actions in this case is given by
\begin{eqnarray}
(\hat{L}(\eta_{I}))^{2}|_{1}h_{\alpha}[0,1]_{A}\UI{B}
&\!\!\!=\!\!\!&-\frac{3}{16}h_{\alpha}[0,1]_{A}\UI{B},
\label{eq:basiclength} \\
(\hat{L}(\eta_{I}))^{2}|_{2}\left(h_{\alpha}[0,1]_{A}\UI{B}
h_{\alpha}[0,1]_{C}\UI{D}\right)
&\!\!\!=\!\!\!&-\frac{1}{4}\left(h_{\alpha}[0,1]_{A}\UI{D}
h_{\alpha}[0,1]_{C}\UI{B}
-\frac{1}{2}h_{\alpha}[0,1]_{A}\UI{B}
h_{\alpha}[0,1]_{C}\UI{D}\right).
\nonumber
\end{eqnarray}

Eq.(\ref{eq:basiclength}) tells us that,
while each tensor product of $l$
spinor propagators
$$
h_{\alpha}[0,1]_{A_{1}\cdots A_{l}}\UI{B_{1}\cdots B_{l}}\equiv
\prod_{i=1}^{l}h_{\alpha}[0,1]_{A_{i}}\UI{B_{i}}
$$
are not eigenvectors unless $l=1$ or $0$,
the symmetrized tensor products yield the eigenvectors. Using
eq.(\ref{eq:lengthsepa}), we can easily calculate their eigenvalues:
\begin{eqnarray}
(\hat{L}(\eta_{I}))^{2}h_{\alpha}
[0,1]_{A_{1}\cdots A_{l}}\UI{(B_{1}\cdots B_{l})}
&=& p\left(-\frac{3}{16}\right)
h_{\alpha}[0,1]_{A_{1}\cdots A_{l}}\UI{(B_{1}\cdots B_{l})}
\nonumber \\
& &-\frac{1}{4}\sum_{1\leq i<j\leq l}\left(
h_{\alpha}[0,1]_{A_{1}\cdots A_{i}\cdots A_{j}\cdots A_{l}}
\UI{(B_{1}\cdots B_{j}\cdots B_{i}\cdots B_{l})}-\frac{1}{2}
h_{\alpha}[0,1]_{A_{1}\cdots A_{l}}\UI{(B_{1}\cdots B_{l})}\right)
\nonumber \\
&=&(-\frac{3p}{16}-\frac{1}{8}\cdot\frac{p(p-1)}{2})
h_{\alpha}[0,1]_{A_{1}\cdots A_{l}}\UI{(B_{1}\cdots B_{l})}
\nonumber \\
&=&-\frac{p(p+2)}{16}
h_{\alpha}[0,1]_{A_{1}\cdots A_{l}}\UI{(B_{1}\cdots B_{l})}.
\label{eq:evlength}
\end{eqnarray}
The result is problematic because the eigenvalue is negative
while classically $(L(\eta_{I}))^{2}$ takes positive values\footnote{
This situation arises also in (3+1)-dimensions, where the
\lq squared-infinitesimal area'
operator which is naively defined from the classical expression has
negative eigenvalues on the spin network states.
This issue might tell us that we must consider
more \lq singular' configurations in which loops are non-analytic,
i. e. have kinks or intersections, almost everywhere.}.
If we ignore the minus sign by taking the \lq absolute value'
of this operator, however, we see that the combinatorial solution
$\Psi^{combi}_{\{(\alpha_{i},l_{i})\}}$ becomes the eigenvector
of the length operator $\hat{L}(\eta)$ if
$\eta$ does not pass through the
intersections. The eigenvalue in this case is given by
\beq
\frac{1}{4}\sum_{i}n(i)\sqrt{l_{i}(l_{i}+1)},
\eeq
where the loop $\alpha_{i}$ is supposed to intersect
with $\eta$ $n(i)$-times. We should mention that
this result is almost identical to that of the \lq area operator'
in (3+1)-dimensions.
This is probably because the 2-dimensional
surface and the 1-dimensional
curve are respectively the dual of the loop in 3-  and 2-dimensions.

%%%%%%%%%%%%%%%%%%%%%%%%%%%%%%%%%%%%%%%%%%%%%%%%%%%%%%%%%%%%%%%%%%%%

\section{Discussion:
an extension to (3+1)-dimensional quantum gravity}

In this paper we have investigated the Dirac quantization
of (2+1)-dimensional analog of Ashtekar's
approach to quantum gravity
using the spin network states.
We have regularized the Hamiltonian constraint by introducing the
(local) curvilinear coordinate frame
in which the loop parameters play the role
of coordinates. Then we have found a set of combinatorial solutions
$$
\{\Psi^{combi.}_{\{(\alpha_{i},l_{i})\}}\}
$$
to the Hamiltonian constraint. Each solution is labelled by a set
$\{\alpha_{i}\}$ of smooth curves
which may intersect with one another or
with themselves and which is equipped with
a non-negative number $l_{i}$.
These combinatorial solutions
correspond to the classical solutions
which do not belong to the solution space of Witten's formalism
and in which the induced metric on $\Sigma$ is degenerate.
We can construct the operator
which measures the area of a 2-dimensional
region. The result of its action depends on the way of regularization
and the action on the combinatorial solutions
in a more natural regularization
yields nonvanishing result.
This can be considered as a quantum effect.

As a by-product we have established a prescription for calculating
the action of an operator $\hat{O}^{(n)}$
which involves an $n$-th order product
of the momentum $\tE^{i}_{a}$. We first work out all the actions
on the basic configurations which consist of at most $n$ parallel
propagators in the spinor representation.
Next we divide the action of the operator as:
$$
\hat{O}^{(n)}=\sum_{k=1}^{n}\hat{O}^{(n)}|_{k},
$$
where $\hat{O}^{(n)}|_{k}$ denotes the action of $\hat{O}^{(n)}$
on the sets each of which is composed of $k$ propagators.
The problem of calculating the action of $\hat{O}^{(n)}$ thus
reduces to that of combinatorics. In particular, the graphical
representation introduced in this paper is expected to provide
a powerful tool for calculation.

In this paper we did not exploit the
merit of (2+1)-dimensions, except
the possibility of the naive regularization.
Thus we anticipate that
most of the results obtained in this paper can be extended
to (3+1)-dimensions.

An essential difference between these two cases is that we cannot
naively apply to (3+1)-dimensions
the regularization of the Hamiltonian
constraint adopted in this paper.
Roughly speaking, many attempts to
regularize the Hamiltonian constraint in
(3+1)-dimensions have been made in two directions.
One is the point splitting
regularization \cite{smol} \cite{bren} \cite{bori} and the other is
the regularization based on the extended loops\cite{bart}.
The former needs to be accompanied with
a multiplicative renormalization and the latter is
an attempt to elaborate the
\lq flux-tube regularization' introduced in
ref.\cite{jacob}.
If we naively use the
point-splitting regularization to (3+1)-dimensions,
$\utep_{ij}\dot{\alpha}^{i}(s)\dot{\beta}^{j}(t)$ and
$\tB^{a}\equiv\frac{1}{2}\otep^{ij}F^{a}_{ij}$
which appeared in (2+1)-dimensions is replaced, respectively,  by
\footnote{We have used here
the conventional notation in (3+1)-dimensions.}
$\utep_{abc}\dot{\alpha}^{b}(s)\dot{\beta}^{c}(t)$ and
$\tB^{ia}\equiv\frac{1}{2}\otep^{abc}F^{i}_{bc}$.
Naively we cannot regularize
the Hamiltonian constraint in a way
which respects both diffeomorphism
covariance and invariance under the reparametrization of the curves.
Nevertheless we can consider that, if a modification is made,
the combinatorial solutions
$\Psi^{combi.}_{\{(\alpha_{i},\l_{i})\}}$
obtained in this paper still
yield the solutions of the renormalized
Hamiltonian constraint ${\cal H}^{ren}$
in (3+1)-dimensions. If we make the replacement mentioned above,
the basic action of ${\cal H}^{ren}$ is represented in the
same way as that shown in Appendix, up to constant factors which
may depend on the angle $\theta$ between
two curves measured in some fixed
background metric.
First we should note that, in the configurations considered here,
the \lq acceleration term', if any, cannot be distinguished from
a diffeomorphism\footnote{
In other words, the acceleration terms disappear if we adopt the
regularization similar to that used in this paper.}.
We can thus neglect this term.

In obtaining the combinatorial solutions, the overall
factor is irrelevant.
We suppose that the ratio of the factor in (A.3)
to that in (A.2) be $Z(\theta)$
and that the ratio of the factor in (A.4)
to that in (A.2) be $Y(\theta)$. The discussion analogous to that
leading to eq.(\ref{eq:combisol}) shows that,
if we use the intertwining
operator:
\beq
I^{\star}_{(Y,Z)}(m,n)\equiv\sum_{r=0}^{\min(m,n)}
\left(\prod_{i=0}^{r}
\frac{-Z(\theta)}{(m+n-2i)Y(\theta)+1}\right)I^{r}(m,n)
\eeq
instead of $I^{\circ}(m,n)$ in eq.(\ref{eq:intertwiner}), then
$\Psi^{combi.\star}_{\{(\alpha_{i},l_{i})\}}$ thus obtained
becomes the combinatorial solution to
the renormalized Hamiltonian constraint
$$
{\cal H}^{ren}\Psi=0.
$$
In (3+1)-dimensions, however,
we also have to consider \lq essentially
3 dimensional intersections' at which at least three curves
with non-degenerate tangent vectors intersect.
These intersections may
yield the combinatorial solutions
which is physically more interesting.
While the calculation may become more
complicated than in (2+1)-dimensions,
we believe that no essential difficulty arises in carrying out
this calculus.

We can easily see that the combinatorial
solutions $\Psi^{combi.\star}$
in (3+1)-dimensions also correspond to classical solutions
whose metric is degenerate. Hence naively the action of
the volume operator on these solutions is expected to vanish.
As we have seen in \S 5, however, the action of an operator is
regularization-dependent. In particular, it is suspicious
whether there exists a prescription
in which both the Hamiltonian
constraint and the volume operator are regularized consistently.
Thus it is no wonder that the combinatorial solutions should
have non-vanishing action of the volume operator.

In this paper we have not exploited
the essential virtue of (2+1)-gravity,
namely, its topological nature. The analysis made here gave
only results which is somewhat formal and abstract.
To make further study we need to acquire an intimate acquaintance
with the classical solutions in a fixed topology.
By comparing Dirac's quantization method in a fixed topology
to the reduced phase space quantization which is based on the space
of classical solutions, we may have some conceptually important
lessons on the quantum gravity in (3+1)-dimensions.
While we expect that Dirac's quantization yields physical
states corresponding to quantum fluctuations
from the classical solutions\cite{ashte2},
we have not yet found such solutions.
These solutions as well as the solutions in (3+1)-dimensions
which is the quantum counterpart of the classical solutions
with non-degenerate metric may play some important roles.
While we have not mentioned here, the closure of the constraints
under the commutation relations is essential to the question whether
the Dirac quantization can be carried out consistently.
We have shown in (2+1)-dimensions that the algebra is closed,
up to the commutator of two Hamiltonian constraints.
This remaining commutator is the most interesting in studying
quantum gravity and its detailed investigation is longed for.

\vskip2.5cm

\noindent Acknowledgments

I would like to thank Prof. K. Kikkawa,
Prof. H. Itoyama and H. Kunitomo

for useful discussions and careful readings of the manuscript.
This work is supported by the Japan Society for the
Promotion of Science.

%%%%%%%%%%%%%%%%%%%%%%%%%%%%%%%%%%%%%%%%%%%%%%%%%%%%%%%%%%%%%%%%%%%
\newpage

\appendix

\catcode`\@=11
\def\theequation{\Alph{section}.\arabic{equation}}
%%%%%%%%%%%%%%%%%%%%%
\vspace{1.2in}
\section{Appendix}

\vspace{.8in}
In this appendix we list the action of the Hamiltonian constraint
on the basic configurations. Here the graphical notation is used.
We will denote the parallel propagators along
the curves $\alpha$ and $\beta$ by
an upward vertical arrow and a
rightward horizontal arrow respectively.
For example, we write the tensor product
of the propagators along $\alpha$ and $\beta$ as follows
\footnote{We will
assume that $\alpha$ and $\beta$ intersects once at $x_{0}=
\alpha(s_{0})=\beta(t_{0})$.}:

%%%%%%%%%%%%%%%%%%%%%%%%%%%%%%%%%%%%%%%%
\begin{figure}[h]
\begin{picture}(150,30)(-10,0)
\put(0,0){\makebox(40,30){%
$h_{\alpha}[0,1]_{A}\UI{B}h_{\beta}[0,1]_{C}\UI{D}$}}
\put(40,0){\makebox(20,30){$=$}}
\put(65,0){\usebox{\LBRA}}
\put(90,5){\vector(0,1){20}}
\put(80,15){\vector(1,0){20}}
{\scriptsize
\put(90,2){$A$}
\put(90,25){$B$}
\put(80,15){$C$}
\put(100,15){$D$}}
\put(110,0){\usebox{\RBRA}}
\put(115,0){\makebox(5,30){.}}
\put(120,0){(A.0)}
\end{picture}
\end{figure}
%%%%%%%%%%%%%%%%%%%%%%%%%%%%%%%%%%%%%%

It is convenient to separate $\hat{\cal H}(\tN)$ as
in eq.(\ref{eq:sepa}):
$$
\hat{\cal H}(\tN)=\hat{\cal H}(\tN)_{1}+\hat{\cal H}(\tN)_{2}.
$$

Because we can show that the action on the
smooth lines without intersection vanishes:

%%%%%%%%%%%%%%%%%%%%%%%%%%%%%%%%%%%%%%%%%%%%%%%
\begin{figure}[h]
\begin{picture}(150,30)(-10,0)
\put(0,0){\makebox(15,30){$\hat{\cal H}_{1}(\tN)$}}
\put(15,0){\usebox{\LBRA}}
\put(25,5){\vector(0,1){20}}
{\scriptsize
\put(25,2){$A$}
\put(25,25){$B$}}
\put(35,0){\usebox{\RBRA}}
\put(40,0){\makebox(20,30){$=$}}
\put(60,0){\makebox(15,30){$\hat{\cal H}_{2}(\tN)$}}
\put(75,0){\usebox{\LBRA}}
\put(88,5){\vector(0,1){20}}
\put(92,5){\vector(0,1){20}}
{\scriptsize
\put(88,2){$A$}
\put(88,25){$B$}
\put(92,2){$C$}
\put(92,25){$D$}}
\put(100,0){\usebox{\RBRA}}
\put(105,0){\makebox(15,30){$=0$,}}
\put(120,0){(A.1)}
\end{picture}
\end{figure}
%%%%%%%%%%%%%%%%%%%%%%%%%%%%%%%%%%%%%%
$\quad$\\
we have only to concentrate on the action at the vertex.
In the present case the only vertex
is at $x_{0}$. In order to simplify the notation of the equations,
we introduce the following \lq rescaled Hamiltonian':
$$
(\hat{\cal H}^{\prime}_{x_{0}})_{I}
\equiv-\sigma(\alpha,\beta)\frac{8}{\tN(x_{0})}
\left(\mbox{ the action of
$\hat{\cal H}_{I}(\tN)$ at $x_{0}$}\right),
\quad(I=1,2).
$$

In the following we provide some of
the basic actions of the Hamiltonian constraint
in the graphical notation. The dot in the diagram denotes the
part on which the area derivative $\AD$ acts,
or the location into which
the magnetic field $\tB$ is inserted.
Using equations (A.1-8) and
identities(\ref{eq:identity1})(\ref{eq:2spi2})
(\ref{eq:2spi3}), we can write out all the basic action,
namely all the action on
a single propagator and a pair of propagators, of the Hamiltonian
constraint $\hat{\cal H}(\tN)$.

%%%%%%%%%%%%(A.2)%%%%%%%%%%%%%%%%%%%%%%%%%%%%%%%%
\begin{figure}[h]
\begin{picture}(150,70)(-10,0)
\put(0,40){\makebox(15,30){$\Hp_{1}$}}
\put(15,40){\usebox{\LBRA}}
%%%begin lhs input%%%
\put(30,45){\line(0,1){10}}
\put(30,55){\vector(1,0){10}}
{\scriptsize
\put(30,42){$A$}
\put(40,55){$B$}}
%%%end lhs input%%%

\put(45,40){\usebox{\RBRA}}
\put(50,40){\makebox(10,30){$=$}}
\put(60,40){\usebox{\LBRA}}

%%%begin rhs input%%%%
\put(80,45){\line(0,1){10}}
\put(80,55){\vector(1,0){10}}
\put(80,55){\circle*{1.5}}
{\scriptsize
\put(80,42){$A$}
\put(90,55){$B$}}

%%%end rhs input%%%
\put(110,40){\usebox{\RBRA}}
\put(115,40){\makebox(5,30){,}}
\put(120,40){(A.2)}

%%%%%%%%%%%%%%%(A.3)%%%%%%%%%%%%%%%%%%%%%%%%%%%%%%%%%%%

\put(0,0){\makebox(15,30){$\Hp_{2}$}}
\put(15,0){\usebox{\LBRA}}
%%%begin lhs input%%%
\put(30,5){\vector(0,1){20}}
\put(20,15){\vector(1,0){20}}
{\scriptsize
\put(30,2){$A$}
\put(30,25){$B$}
\put(18,15){$C$}
\put(40,15){$D$}}
%%%end lhs input%%%

\put(40,0){\usebox{\RBRA}}
\put(45,0){\makebox(10,30){$=$ $2$}}
\put(55,0){\usebox{\LBRA}}

%%%begin rhs input%%%%
\put(60,15){\line(1,0){10}}
\put(71,14){\vector(1,0){9}}
\put(70,15){\vector(0,1){10}}
\put(71,5){\line(0,1){9}}
\put(71,14){\circle*{1.5}}
{\scriptsize
\put(70,2){$A$}
\put(70,25){$B$}
\put(58,15){$C$}
\put(80,14){$D$}}
\put(80,0){\makebox(15,30){$-$}}
\put(95,16){\line(1,0){9}}
\put(105,15){\vector(1,0){10}}
\put(104,16){\vector(0,1){9}}
\put(105,5){\line(0,1){10}}
\put(104,16){\circle*{1.5}}
{\scriptsize
\put(105,2){$A$}
\put(105,25){$B$}
\put(93,16){$C$}
\put(115,15){$D$}}

%%%end rhs input%%%
\put(115,0){\usebox{\RBRA}}
\put(120,0){\makebox(5,30){,}}
\put(125,0){(A.3)}
\end{picture}
\end{figure}

%%%%%%%%%%%%(A.4)%%%%%%%%%%%%%%%%%%%%%%%%%%%%%%%%%%%%%%%%%
\begin{figure}[h]
\begin{picture}(150,70)(-10,0)
\put(0,40){\makebox(15,30){$\Hp_{2}$}}
\put(15,40){\usebox{\LBRA}}
%%%begin lhs input%%%
\put(27,45){\vector(0,1){20}}
\put(30,45){\line(0,1){10}}
\put(30,55){\vector(1,0){10}}
{\scriptsize
\put(27,42){$A$}
\put(27,65){$B$}
\put(30,42){$C$}
\put(40,55){$D$}}
%%%end lhs input%%%

\put(40,40){\usebox{\RBRA}}
\put(45,40){\makebox(10,30){$=$}}
\put(55,40){\usebox{\LBRA}}

%%%begin rhs input%%%%
\put(66,55){\vector(0,1){10}}
\put(70,55){\vector(1,0){10}}
\put(70,45){\line(0,1){6}}
\put(66,45){\line(0,1){6}}
\put(66,51){\line(1,1){4}}
\put(66,55){\line(1,-1){4}}
\put(70,55){\circle*{1.5}}
{\scriptsize
\put(66,42){$A$}
\put(66,65){$B$}
\put(70,42){$C$}
\put(80,55){$D$}}
\put(80,40){\makebox(15,30){$-$}}
\put(101,55){\vector(0,1){10}}
\put(105,55){\vector(1,0){10}}
\put(105,45){\line(0,1){6}}
\put(101,45){\line(0,1){6}}
\put(101,51){\line(1,1){4}}
\put(101,55){\line(1,-1){4}}
\put(101,55){\circle*{1.5}}
{\scriptsize
\put(101,42){$A$}
\put(101,65){$B$}
\put(105,42){$C$}
\put(115,55){$D$}}

%%%end rhs input%%%
\put(115,40){\usebox{\RBRA}}
\put(120,40){\makebox(5,30){,}}
\put(125,40){(A.4)}

%%%%%%%%%%%%%%%(A.5)%%%%%%%%%%%%%%%%%%%%%%%%%%%%%%%%%%%%%%

\put(0,0){\makebox(15,30){$\Hp_{2}$}}
\put(15,0){\usebox{\LBRA}}
%%%begin lhs input%%%
\put(30,5){\vector(0,1){20}}
\put(27,15){\vector(0,1){10}}
\put(20,15){\line(1,0){7}}
{\scriptsize
\put(20,15){$A$}
\put(27,25){$B$}
\put(30,2){$C$}
\put(30,25){$D$}}
%%%end lhs input%%%

\put(40,0){\usebox{\RBRA}}
\put(45,0){\makebox(10,30){$=$}}
\put(55,0){\usebox{\LBRA}}

%%%begin rhs input%%%%
\put(66,19){\vector(0,1){6}}
\put(60,15){\line(1,0){6}}
\put(70,5){\line(0,1){10}}
\put(70,19){\vector(0,1){6}}
\put(66,15){\line(1,1){4}}
\put(66,19){\line(1,-1){4}}
\put(66,19){\circle*{1.5}}
{\scriptsize
\put(60,15){$A$}
\put(66,25){$B$}
\put(70,2){$C$}
\put(70,25){$D$}}
\put(80,0){\makebox(15,30){$-$}}
\put(101,19){\vector(0,1){6}}
\put(95,15){\line(1,0){6}}
\put(105,5){\line(0,1){10}}
\put(105,19){\vector(0,1){6}}
\put(101,15){\line(1,1){4}}
\put(101,19){\line(1,-1){4}}
\put(105,19){\circle*{1.5}}
{\scriptsize
\put(95,15){$A$}
\put(101,25){$B$}
\put(105,2){$C$}
\put(105,25){$D$}}

%%%end rhs input%%%
\put(115,0){\usebox{\RBRA}}
\put(120,0){\makebox(5,30){,}}
\put(125,0){(A.5)}
\end{picture}
\end{figure}

%%%%%%%%%%%%(A.6)%%%%%%%%%%%%%%%%%%%%%%%%%%%%%%%%%%%%%%%%%
\begin{figure}[t]
\begin{picture}(150,70)(-10,0)
\put(0,40){\makebox(15,30){$\Hp_{2}$}}
\put(15,40){\usebox{\LBRA}}
%%%begin lhs input%%%
\put(32,55){\vector(1,0){8}}
\put(28,55){\vector(-1,0){8}}
\put(28,45){\line(0,1){10}}
\put(32,45){\line(0,1){10}}
{\scriptsize
\put(28,42){$A$}
\put(20,55){$B$}
\put(32,42){$C$}
\put(40,55){$D$}}
%%%end lhs input%%%

\put(40,40){\usebox{\RBRA}}
\put(45,40){\makebox(10,30){$=$}}
\put(55,40){\usebox{\LBRA}}

%%%begin rhs input%%%%
\put(72,55){\vector(1,0){8}}
\put(68,55){\vector(-1,0){8}}
\put(68,45){\line(0,1){6}}
\put(72,45){\line(0,1){6}}
\put(68,51){\line(1,1){4}}
\put(68,55){\line(1,-1){4}}
\put(72,55){\circle*{1.5}}
{\scriptsize
\put(68,42){$A$}
\put(60,55){$B$}
\put(72,42){$C$}
\put(80,55){$D$}}
\put(80,40){\makebox(15,30){$-$}}
\put(107,55){\vector(1,0){8}}
\put(103,55){\vector(-1,0){8}}
\put(103,45){\line(0,1){6}}
\put(107,45){\line(0,1){6}}
\put(103,51){\line(1,1){4}}
\put(103,55){\line(1,-1){4}}
\put(103,55){\circle*{1.5}}
{\scriptsize
\put(103,42){$A$}
\put(95,55){$B$}
\put(107,42){$C$}
\put(115,55){$D$}}

%%%end rhs input%%%
\put(115,40){\usebox{\RBRA}}
\put(120,40){\makebox(5,30){,}}
\put(125,40){(A.6)}

%%%%%%%%%%%%%%(A.7)%%%%%%%%%%%%%%%%%%%%%%%%%%%%%%%%%%%%%%%

\put(0,0){\makebox(15,30){$\Hp_{2}$}}
\put(15,0){\usebox{\LBRA}}
%%%begin lhs input%%%
\put(32,15){\line(1,0){8}}
\put(20,15){\line(1,0){8}}
\put(28,15){\vector(0,1){10}}
\put(32,15){\vector(0,1){10}}
{\scriptsize
\put(28,25){$B$}
\put(20,15){$A$}
\put(32,25){$D$}
\put(40,15){$C$}}
%%%end lhs input%%%

\put(40,0){\usebox{\RBRA}}
\put(45,0){\makebox(10,30){$=$}}
\put(55,0){\usebox{\LBRA}}

%%%begin rhs input%%%%
\put(72,15){\line(1,0){8}}
\put(60,15){\line(1,0){8}}
\put(68,19){\vector(0,1){6}}
\put(72,19){\vector(0,1){6}}
\put(68,15){\line(1,1){4}}
\put(68,19){\line(1,-1){4}}
\put(72,15){\circle*{1.5}}
{\scriptsize
\put(68,25){$B$}
\put(60,15){$A$}
\put(72,25){$D$}
\put(80,15){$C$}}
\put(80,0){\makebox(15,30){$-$}}
\put(107,15){\line(1,0){8}}
\put(95,15){\line(1,0){8}}
\put(103,19){\vector(0,1){6}}
\put(107,19){\vector(0,1){6}}
\put(103,15){\line(1,1){4}}
\put(103,19){\line(1,-1){4}}
\put(103,15){\circle*{1.5}}
{\scriptsize
\put(103,25){$B$}
\put(95,15){$A$}
\put(107,25){$D$}
\put(115,15){$C$}}

%%%end rhs input%%%
\put(115,0){\usebox{\RBRA}}
\put(120,0){\makebox(5,30){,}}
\put(125,0){(A.7)}
\end{picture}
\end{figure}

%%%%%%%%%%%%%%%(A.8)%%%%%%%%%%%%%%%%%%%%%%%%%%%%%%%%%%%%%%

\begin{figure}[h]
\begin{picture}(150,30)(-10,0)
\put(0,0){\makebox(15,30){$\Hp_{2}$}}
\put(15,0){\usebox{\LBRA}}
%%%begin lhs input%%%
\put(31,14){\vector(1,0){9}}
\put(20,16){\line(1,0){9}}
\put(29,16){\vector(0,1){9}}
\put(31,5){\line(0,1){9}}
{\scriptsize
\put(29,25){$B$}
\put(20,16){$C$}
\put(31,2){$A$}
\put(40,14){$D$}}
%%%end lhs input%%%

\put(40,0){\usebox{\RBRA}}
\put(45,0){\makebox(10,30){$=$}}
\put(55,0){\makebox(15,30){$\Hp_{2}$}}
\put(70,0){\usebox{\LBRA}}

%%%begin rhs input%%%%
\put(78,17){\vector(1,0){12}}
\put(81,14){\vector(1,0){9}}
\put(78,5){\line(0,1){12}}
\put(81,5){\line(0,1){9}}
{\scriptsize
\put(90,17){$B$}
\put(78,2){$A$}
\put(90,14){$D$}
\put(81,2){$C$}}
\put(95,0){\usebox{\RBRA}}
%%%end rhs input%%%
\put(100,0){\makebox(20,30){$=0$ .}}
\put(125,0){(A.8)}
\end{picture}
\end{figure}

%%%%%%%%%%%%%%%%%%%%%%%%%%%%%%%%%%%%%%%%%%%%%%%%%%%%%

%%%%%%%%%%%%%%%%%%%%%%%%%%%%%%%%%%%%%%%%%%%%%%%%%%%%%%%%%%%%%%%%%%

\end{document}